\documentclass[]{aa}  
\usepackage{txfonts}
\usepackage{geometry}
\usepackage{color}

\newcommand{\arsco}{AR\,Sco}

\begin{document}

   \title{VLA Radio Observations of AR Sco}

   \author{E.~R.~Stanway%
          \inst{1},
          T.~R.~Marsh%
          \inst{1},
          P.~Chote\inst{1},
          B.~{T.}~G{\"a}nsicke\inst{1},
          D.~Steeghs\inst{1}
          \and
          P.~J~Wheatley\inst{1}
          }

   \institute{Department of Physics, University of Warwick,
              Gibbet Hill Road, Coventry, CV4 7AL, UK\\
              \email{e.r.stanway@warwick.ac.uk}
             }

 \authorrunning{E.~R.~Stanway et al.}

   \date{Received 2017 Nov 29; accepted 2018 Jan 22}

  \abstract
  % context heading (optional)
    {} 
  % aims heading (mandatory)
    {AR Scorpii is unique amongst known white dwarf binaries in showing powerful pulsations extending to radio frequencies. Here we aim to investigate the multi-frequency radio emission of AR Sco in detail, in order to constrain its origin and emission mechanisms.}
  % methods heading (mandatory)
    {We present interferometric radio frequency imaging of AR Sco at 1.5, 5 and 9\,GHz, analysing the total flux and polarization behaviour of this source at high time resolution (10, 3 and 3\,s), across a full 3.6\,hr orbital period in each band.}
  % results heading (mandatory)
    {We find strong modulation of the radio flux on the orbital period and the orbital sideband of the white dwarf's spin period (also known as the "beat" period). This indicates that, like the optical flux, the radio flux arises predominantly from on or near the inner surface of the M-dwarf companion star.
    The beat-phase pulsations of AR Sco decrease in strength with decreasing frequency. They are strongest at 9\,GHz and at an orbital phase $\sim$0.5.
    Unlike the optical emission from this source, radio emission from AR Sco shows weak linear polarization but very strong circular polarization, reaching $\sim$30\% at an orbital phase $\sim$0.8. We infer the probable existence of a non-relativistic cyclotron emission component, which dominates at low radio frequencies. Given the required magnetic fields, this also likely arises from on or near the M-dwarf.}
    % conclusions heading (optional)
    {}
    

   \maketitle
%
%-------------------------------------------------------------------

\section{Introduction}

AR~Scorpii (hereafter \arsco) is a white dwarf / M dwarf close binary with an orbital period of 3.6\,hr \citep{2016Natur.537..374M}. It shows strong pulsations in brightness associated with the 117.1\,s spin period of its white dwarf, which, uniquely, are seen from the ultraviolet all the way to radio frequencies. The mismatch between spin and orbital periods suggest a relation with the intermediate polar class of white dwarf binaries in which an accreting magnetic white dwarf spins faster than the binary orbit due to accretion from a disc that is disrupted close to the white dwarf, where the magnetic field dominates \citep[e.g.][]{1982MNRAS.198..975W}. However, the source shows little or no evidence for an accretion disc which might suggest a more natural match to the asynchronous polars, in which the difference between spin and orbital periods is likely attributable to an impulse from a nova event \citep{1988ApJ...332..282S}. On the other hand, asynchronous polars typically present an orbit-spin period difference of less than 1\% \citep{2002AIPC..637....3W}, and \arsco\ is far from meeting that criterion. In fact, \arsco\ is distinct from all known white dwarf / main sequence binary systems in a number of key respects, but primarily in the lack of any evidence for accretion and the great strength of its pulsations which are almost 100\% modulated at ultraviolet wavelengths and which extend all the way to radio wavelengths.

\citeauthor{2016Natur.537..374M} identified three dominant components in the frequency spectrum of \arsco's flux variability: the orbital period of the binary around its barycentre (3.56 hours), the spin period of the white dwarf (117.1\,s) and the beat period between these (118.2\,s). Given the strength of the beat period, and the power-law spectral energy distribution extending from the optical to the radio, it was proposed that the source flux was dominated by synchrotron emission, arising from an interaction between the two components of the binary. Further, the lack of clear signatures of an accretion disk component in the spectrum, together with the energy requirements of the system, suggested that this emission was powered by energy liberated by the spin down of a magnetic, rapidly-rotating white dwarf.

\arsco\ has since been the subject of intensive investigation and theoretical analysis. \citet{2017ApJ...845L...7L}, using optical data from Kepler and CRTS, refined the orbital parameters and demonstrated that while the orbital waveform is stable on timescales of $\sim78$\,days (the duration of the \textit{K2} campaign), it alters slowly over timescales on the order of years, while also showing aperiodic variations superimposed on the regular periodic flux changes. \citet{2017A&A...601L...7M} presented observations taken at 8.5\,GHz with the Australian Long Baseline Array (LBA) which confirmed that the radio emission originates from a compact point source, with no evidence for extended radio jets. \citet{2017NatAs...1E..29B} obtained polarimetric data in the optical, identifying very strong (40\%) spin-modulated linear polarization, and a few percent circular polarization, which also varies with time. These observations strengthen the case for interpretation of \arsco\ as the first-known white dwarf pulsar, with $\sim6$\% of the spin-down power reprocessed by magnetospheric interactions to generate a self-absorbed synchrotron spectrum that dominates from the radio to the optical. 

The initial model proposed by \citet{2016Natur.537..374M} was of a rapidly rotating white dwarf, whose bipolar magnetic field `whips' past the tidally-locked red dwarf twice in every spin exciting emission from a hotspot on the face of the M-dwarf directed towards the white dwarf. The binary orbital motion then leads to a dominant modulation on the beat period. \citet{2016ApJ...831L..10G} developed this model further, identifiying \arsco\ as a near-perpendicular rotator whose open field lines sweep through the M-dwarf's stellar wind, accelerating electrons in a bow shock region above the M-dwarf surface. They note that, in their near edge-on model, the hemisphere seen by the observer has magnetic field directions which cancel out, suggesting that little circular polarization should be seen. 
\citet{2017ApJ...835..150K} has also developed a model for this system, suggesting that the white dwarf spin-down energy is instead dissipated through magnetic reconnection in the M-dwarf atmosphere, dubbing this process synchronization, and the system as a whole a {\it synchronar}. In this model \arsco\ occupies a short-lived, transitional state lying between intermediate polars (which contain a rapidly rotating white dwarf in a binary with a tidally-locked red dwarf)
and traditional polars (in which the white dwarf spin is magnetically locked to the orbital period: the field lines of the white dwarf interact with the field of the companion, and force the white dwarf into synchronous rotation). This model naturally accounts for shifts of the flux maximum away from mid-orbital phase due to precession of the magnetic poles, and for rapid flaring due to magnetic storms in the M-dwarf atmosphere.  Given the complex and challenging task of modelling this source, additional data are useful for refining the interpretation.

While the radio observations presented by \citet{2016Natur.537..374M} confirmed the unusual nature of this source, they were limited to a single hour of data at low spatial resolution, and so could not constrain the orbital modulation in the radio. Similarly, those taken with the VLBI and presented by \citet{2017A&A...601L...7M} confirmed \arsco's identity as a point source and were able to recover an orbital variation lightcurve, but had neither the time resolution nor the frequency range to fully explore the properties of this source in the radio. 

In this paper we present high time resolution radio data obtained with the Karl G. Jansky Very Large Array (VLA) at 1\, 5 and 9\,GHz (L, C and X bands respectively), and use this to explore the properties of \arsco\ on different timescales. We present our data acquisition and reduction in section \ref{sec:obs}. In section \ref{sec:orbit} we consider the radio properties of \arsco\ on timescales of the orbital period, while in section \ref{sec:beat} we identify and investigate pulsations on the system beat period. In section \ref{sec:pol} we investigate evidence for polarization in the source.  In section \ref{sec:disc} we discuss and interpret the radio properties of \arsco, before presenting our conclusions in section \ref{sec:conc}.

%-------------------------------------------------------------------

\section{Data Acquisition and Reduction}\label{sec:obs}

\subsection{Radio Observations}
Radio observations of AR Scorpii were obtained at the Karl G. Jansky Very Large Array (VLA) between 2016 April 28 and 2016 May 02\footnote{Observations associated with programme VLA/16A-338, PI: Marsh.}. Data were taken at frequencies centred around 1.5\,GHz (L band, 6 hours), 5\,GHz (C band, 4 hours) and 9\,GHz (X band, 4 hours), where the integration length was chosen to capture a full 3.6\,hr orbital period (see table \ref{tab:obstable}). Observations were taken with the phase centre at the source location.
At 5\& 9\,GHz, the visibilities were read out every 3 seconds, while at 1.5\,GHz a readout interval of 1\,s was selected. All observations were taken with the telescope in the CnB-configuration. Full polarization information was recorded, and the standard broadband continuum correlator configuration adopted, with a frequency coverage of $1.5\pm0.5$\,GHz, $5.0\pm1.0$\,GHz and $9.0\pm1.0$\,GHz. In the 1.5\,GHz band, two spectral windows (each of 64\,MHz bandwidth) were impossible to calibrate due to strong radio frequency interference (RFI) throughout the calibrator observations, and are omitted from the analysis. In all bands, RFI was flagged as required. Absolute flux and secondary phase calibration were performed using observations of standard calibrators 3C286 and J1626-2951 respectively.

Data were reduced (i.e. automatically flagged, bandpass, gain and flux calibrated)  using the standard VLA data reduction pipeline, integrated in the Common Astronomy Software Applications (CASA, v4.5.2) package. This performs a reduction of continuum data in the Stokes I parameter, calibrating total flux. Further flagging was performed where necessary (particularly in the 1.5\,GHz band) before subsets of the data were imaged, using the {\sc CLEAN} task of CASA. A `Clark' cleaning algorithm was used, with a Briggs weighting parameter of 0.5. Images were constructed using a 1\,arcsecond sampling in the 1.5\,GHz band, and 0.75\,arcsecond sampling in the 5 and 9\,GHz bands.

\begin{figure}
\centering
\includegraphics[width=\columnwidth]{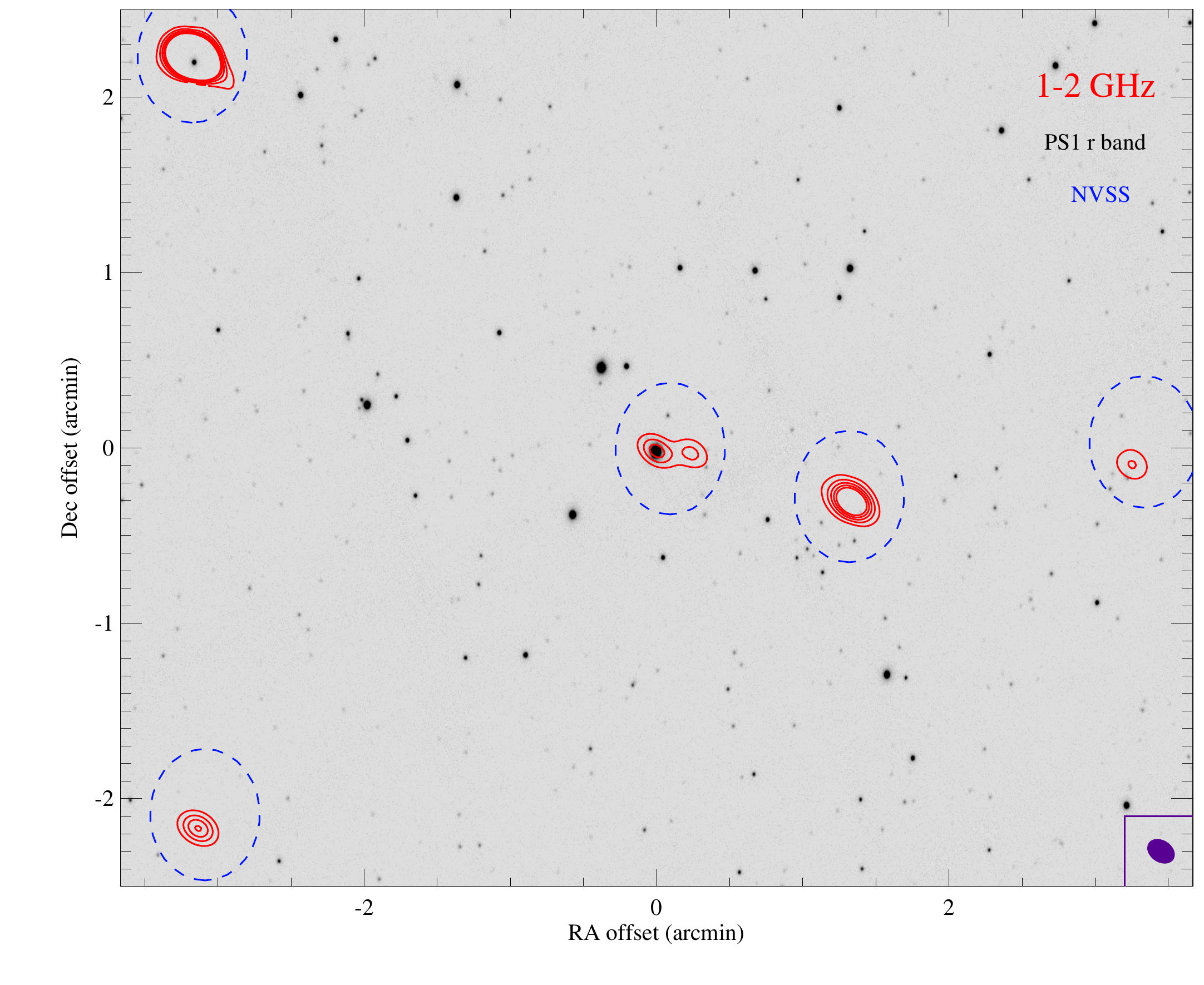}
\caption{The 1.5 GHz radio environment of \arsco. We show our radio contours, for an image constructed from a full orbital period, overlaid on the $r$-band image from the Pan-Starrs PS1 data release. Dashed ellipses show the sources identified in the 1.4\,GHz NRAO VLA Sky Survey (NVSS), at that survey's angular resolution.  The presence of the neighbouring object to \arsco\ (centred in the frame) and its contamination of the NVSS flux reported for this source, is clear. The synthesized beam of our 1.5\,GHz observations is shown in the lower right hand corner. Radio contours indicate multiples of 1.5\,mJy/beam. }
           \label{fig:radio_env}%
\end{figure}

\arsco\ was clearly detected at all bands, even in very short integrations (10\,s at 1.5\,GHz, 3\,s at 5 and 9\,GHz). All bands also show evidence for temporal variability on multiple timescales. Source fluxes were determined using the CASA {\sc IMFIT} command on a region centred at the source location. In the 5 and 9\,GHz bands, the source was straightforwardly fit as a single object. In the 1.5\,GHz band, the flux from \arsco\ is somewhat confused with a neighbouring source at a separation of 14.3\,arcsecond in RA and 1.2\,arcseconds in Declination, which contributed substantially to the archival NVSS flux for \arsco, as figure \ref{fig:radio_env} demonstrates. The source separation is comparable to the synthesized beam of the VLA at this declination and frequency, and so may cause problems, particularly if the primary beam major axis is oriented East-West. Given that each subset of the data has a different $uv$-plane coverage, the severity of this issue varies and is considered on a case by case basis. 
In this band, the two objects were fit simultaneously as point sources. A good fit could be obtained in all cases.

The neighbouring object at RA \& DEC 16$^h$21$^m$46.26$^s$ -22$^0$53$\arcmin$11.86$\arcsec$ (J2000) is consistent with a point source, with no significant temporal variability and a flux of 3.49$\pm$0.08\,mJy at 1.5\,GHz.

We note that the strong variability of \arsco\ can cause problems with image reconstruction from visibility data, and thus that {\sc IMFIT} typically reports the system as being extended in long integrations, despite its point source nature \citep[see][]{2017A&A...601L...7M}. This is particularly true in the 9\,GHz band, and we use the integrated fit properties (i.e. the flux appropriate for an extended source) in this band, while treating \arsco\ as a point source in the other bands. We have checked that this makes no significant difference to flux measurements at 1.5\, and 5\, GHz, but it corrects for the 30\% of flux in the 9\,GHz band which would be lost if peak flux was considered instead.

\begin{table}
\caption{Observation Summary\label{tab:obstable}}
\centering
\begin{tabular}{c c c c}
  \hline\hline
  Date & Observatory & Frequency & Int. Time / s \smallskip\\
  \hline
  2016 Apr 29 & W1m & BG (5500\AA)  & 10,010\\
              &     & Z (8800\AA)  & 9,420
  \smallskip\\
  2016 Apr 29 & VLA & X (9.0\,GHz) & 11,700
  \smallskip\\
  2016 May 01 & VLA & L (1.5\,GHz) & 16,800
  \smallskip\\
  2016 May 02 & VLA & C (5.0\,GHz) & 12,000
  \smallskip\\
  2016 May 04 & W1m & BG (5500\AA) & 4,992\\
              &     & Z (8800\AA) & 7,525\\
  \hline
\end{tabular}
\end{table}

\begin{figure}
\centering
\includegraphics[width=1\columnwidth]{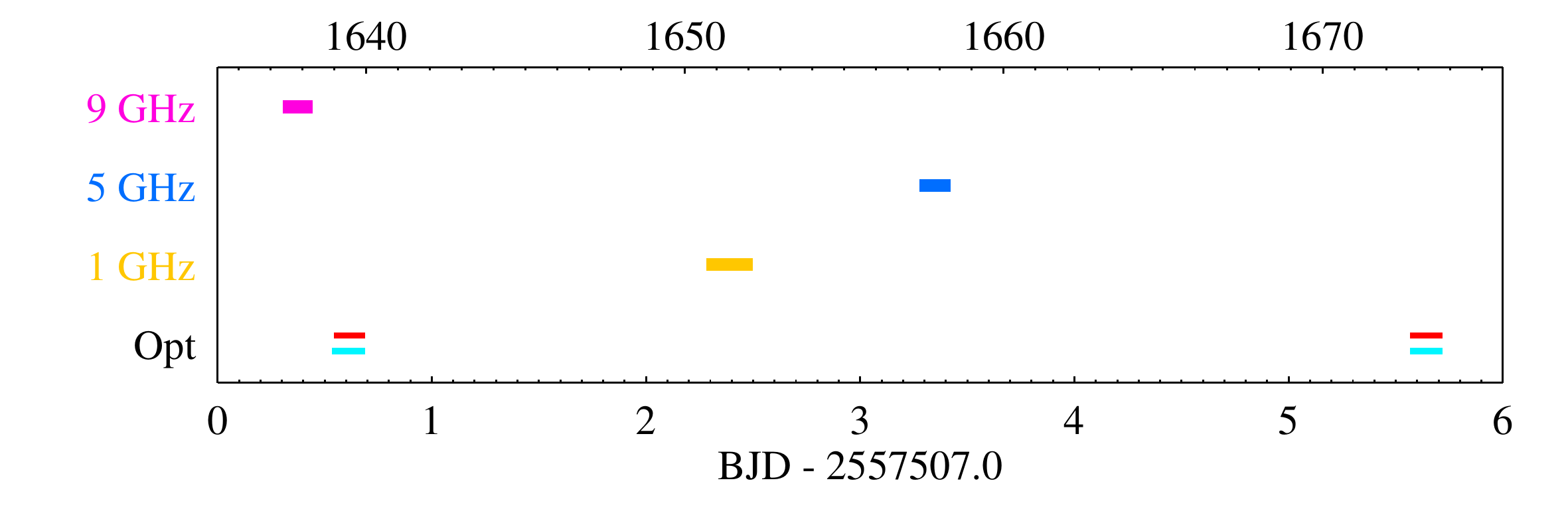}
\caption{Timeline of when observations were taken. Top axis indicates orbital intervals since the zero orbital phase defined in \citet{2016Natur.537..374M}.}
           \label{fig:obsdates}%
\end{figure}

\subsection{Optical Observations}
To improve our constraints on source ephemerides and monitor for unusual behaviour in \arsco, we obtained high time resolution, near-contemporaneous optical observations in two sessions which straddle the dates on which the VLA observed the target, as Figure \ref{fig:obsdates} illustrates. These observations were obtained using the dual-band high speed imager on the Warwick 1m telescope (W1m) on La Palma. Each observation lasted $\sim4$\,hrs, and data were collected simultaneously in blue and red filters with integrations of 10\,s on 2016 Apr 29 and of 2.5 and 5\,s respectively on 2016 May 04. The blue filter has a wide optical bandpass, using BG40 glass, with peak efficiency at $\sim$5500\AA.  The red filter is a $Z$-band, with peak throughput at $\sim$8800\AA. Due to the 3.3\,s detector readout time between integrations, the 10\,s integrations are more efficient in terms of on-sky observing time (see table \ref{tab:obstable}), while the shorter integrations capture more of the dynamic behaviour of this unusual source. Target flux calibration was performed relative to a nearby, non-variable  point source to account for any possible seeing or sky brightness variations.

%-------------------------------------------------------------------

\section{Orbital Period Properties}\label{sec:orbit}

\subsection{SED and spectral slope}

\begin{figure}
\centering
\includegraphics[width=0.95\columnwidth]{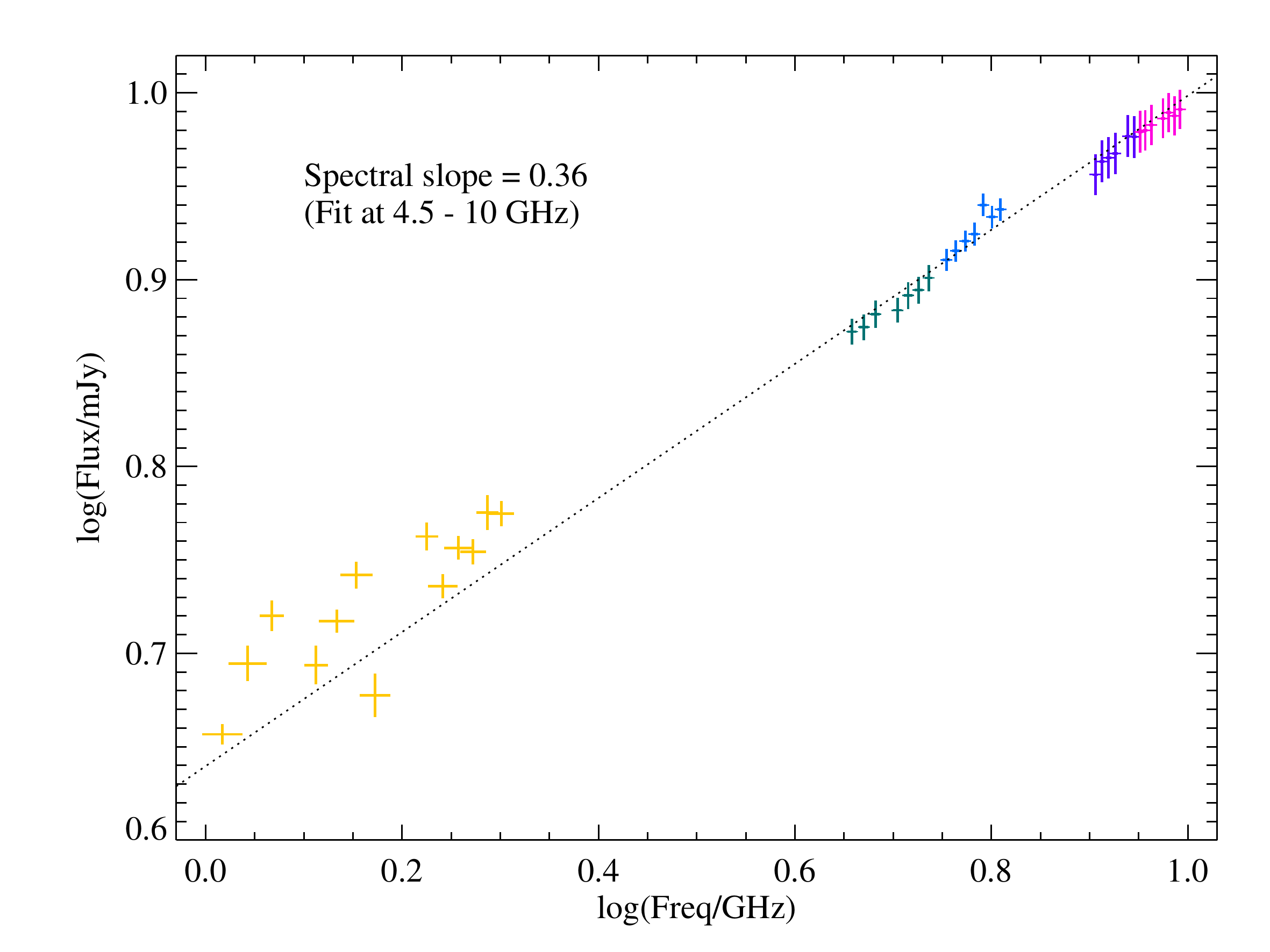}
\caption{Radio spectral energy distribution of \arsco, averaged over a full orbital period. The dashed line indicates a fit in the form $F_\nu \propto \nu^\alpha$ to the spectral slope in the 5 and 9\,GHz bands, extrapolated to the 1.5\,GHz band. The best fit is found at $\alpha=0.358\pm0.015$.}
           \label{fig:orbitsed}%
 \end{figure}

In order to determine the time-averaged spectral energy distribution of \arsco, images were generated from a 3.6\,hr (i.e. full orbital period) interval of observations in each radio band. A secondary phase calibrator was visited every 10-15 minutes, with approximately 2 minutes off-target on each occasion. As a result, the coverage is not fully continuous; the lacunae in the time series are randomly placed with respect to the orbit in each band.

A multifrequency synthesis covering this period was generated for each spectral window (i.e. each covering a bandwidth of 64\,MHz in the 1.5\,GHz band, 128\,MHz in the 5 and 9\,GHz bands), and the flux of \arsco\ in each image determined. In Figure \ref{fig:orbitsed} we show the orbit-averaged spectral energy distribution (SED) of \arsco. Due to the lower signal to noise in the 1.5\,GHz band, and its different pulsation behaviour (see below), we do not use it to constrain the power law fitted to the higher frequency data. The data at 5 and 9\,GHz are consistent with a power law SED with  $F_\nu \propto \nu^\alpha$, where $\alpha=0.358\pm0.015$. While there is a hint that a slightly steeper power law may be appropriate within the 5\,GHz band, the overall fit is good, with an extrapolation of the high frequency power law contributing $>$90\% of the flux in the 1.5 GHz band. The 4-10\,GHz power law fit is similar in spectral slope to the observed flux variation within both the 1.5 and 9\,GHz bands.

\subsection{Orbital  variation}

\begin{figure}
\centering
\includegraphics[width=\columnwidth]{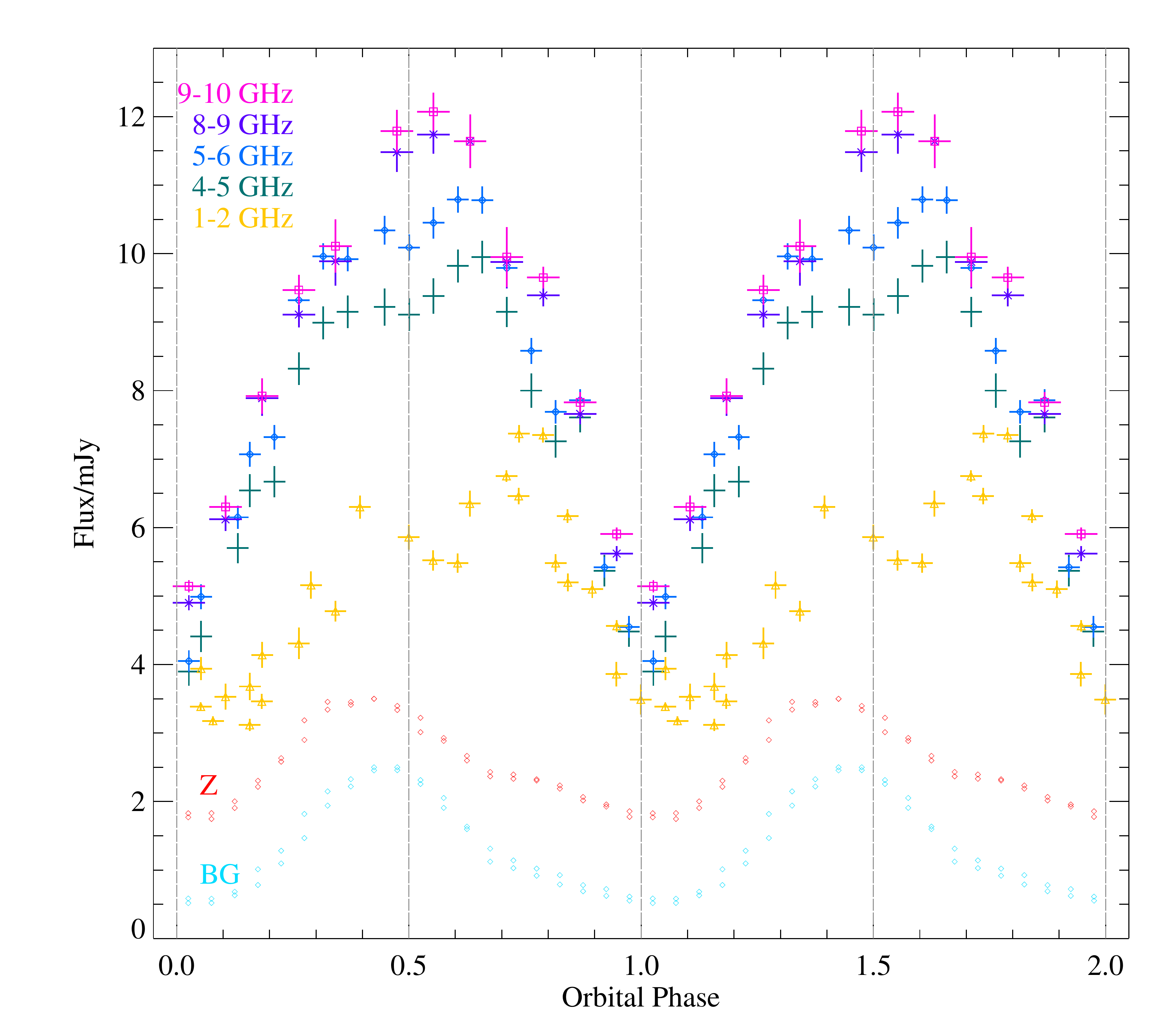}
\caption{Flux variation over the orbital period of \arsco. Data points represents integration over a 10 minute interval (15\,minutes at 9\,GHz), sufficient to average over the two minute beat period flux variation. Results are shown for 1\,GHz bands centred at 1.5, 4.5, 5.5, 8.5 and 9.5\,GHz. The data are phase-folded in the interval between 0 and 1, and repeated between 1 and 2 for clarity. For comparison we also show the W1m optical data as small points in the bottom two lines. The optical data in each observation have been phase folded and the mean in each phase bin calculated; their fluxes are arbitrarily scaled to indicate orbital phase and shape in the optical. Where more than one point occurs at a given frequency and phase, it indicates measurements in different orbits.}  
\label{fig:orbital_flux}%
 \end{figure}

\begin{figure}
\centering
\includegraphics[width=\columnwidth]{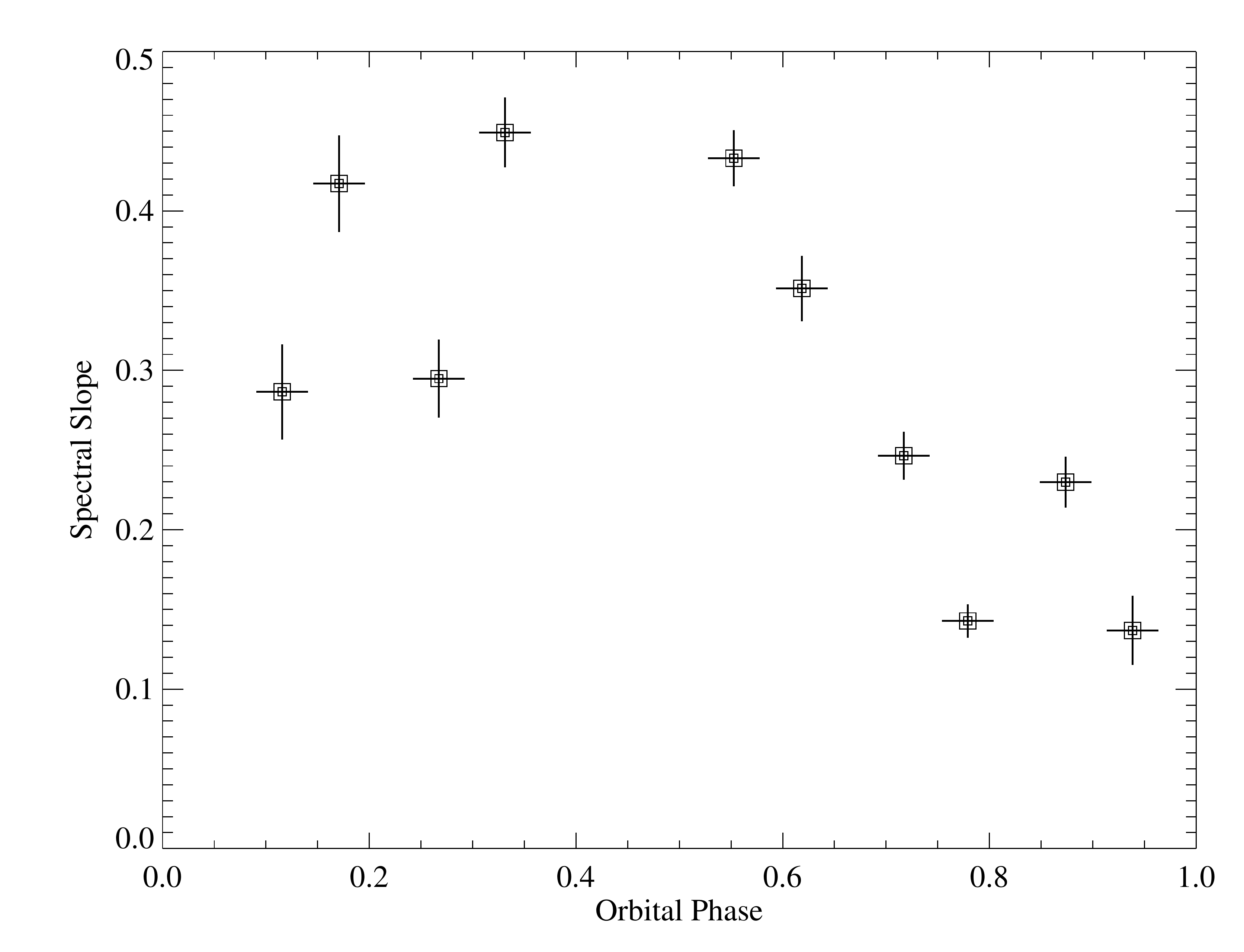}
\caption{Variation in 1-10\,GHz spectral slope over the orbital period of \arsco. Data points represent intervals of 0.05 in orbital phase for which data in all bands is available. In each case, a simple power law fit has been performed to the five bands defined in Figure \ref{fig:orbital_flux}}
           \label{fig:orbital_alpha}%
 \end{figure}

In order to determine flux variability on the orbital period of the system, known from optical observations to be 3.6 hours, we image the field and measure the flux of \arsco\ on intervals corresponding to individual on-source scans, between phase calibrator observations. These have a duration of 10 minutes in the 1.5 and 5\,GHz bands, and 15 minutes in the 9\,GHz band and so are sufficiently long to average over the 118\,s spin{/beat} periods also known from the optical data. We image the data in 1\,GHz bandwidths centred at 1.5, 4.5, 5.5, 8.5 and 9.5\,GHz (i.e. splitting the C and X band observations into high and low frequency subbands).

In Figure \ref{fig:orbital_flux}, we illustrate the frequency dependent total flux variability of \arsco\ with orbital phase.
These measurements are phase folded based on the median phase of each scan, using the zero-phase definition of \citeauthor{2016Natur.537..374M} combined with the updated orbital period based on Kepler observations:
\[ T_0 = 57264.09615 + 0.148533 \times E, \]
where $E$ is an orbit number and the time scale is TDB, corrected to the barycentre of the solar system, expressed as a Modified Julian Day number (BMJD = JD - 2400000.5 + barycentric correction). Zero phase is interpreted as the red dwarf lying closest to the observer along the line of sight. The optical data within each observation have been phase-folded and binned at 0.05 phase intervals with the mean at each phase, in each of the two observing epochs (2016-04-29 and 2016-05-02), plotted in the figure at an arbitrary flux level. The results are shown twice (i.e. repeated at phases between 1 and 2) to better illustrate the overall shape of the orbital flux variation.

As the Figure makes clear, there is considerable variation in the shape of orbital flux variation with frequency. Both the radio data and the optical data reach minimum flux at or just after orbital phase zero. Both are also consistent with small variations in total flux from orbit to orbit, with phase-matched data taken in different orbits not always equal in amplitude. However their behaviour at peak (phase = 0.5) and the overall shape of emission differs. The data at 8-10\,GHz shows a smooth variation in flux from peak to trough, with a sharply defined minimum and a much broader maximum in the light curve. The 4-6\,GHz data shows a smaller differential between the width of maximum and minimum, but a secondary dip in the flux, overlying the peak and occurring just before it in the lightcurve, appears and grows in strength as the frequency drops.  The result is a double peaked light curve, with the first peak (at a phase of $\sim0.35$) weaker than the second (at phase  $\sim0.65$). In the 1.5\,GHz band this behaviour is more pronounced, with the bottom of the light curve much broader and the peak heavily suppressed to leave a double-peaked curve. 

As a result, the 1-10\,GHz spectral index of \arsco, assuming a simple power law, also varies with orbital phase as Figure \ref{fig:orbital_alpha} illustrates. Only those phases with data in all five subbands are fitted and shown. The spectral index decreases with increasing orbital phase, peaking at $\alpha\sim0.5$ at an orbital phase of 0.5 (peak flux), before falling to $\sim$0.1 at a phase of 1 (minimum flux).

The fractional flux variation on the orbital period, defined as $(f_\mathrm{max}-f_\mathrm{min})/(f_\mathrm{max}+f_\mathrm{min})$, is 0.41, 0.44, 0.45, 0.41 and 0.40 at 1.5, 4.5, 5.5, 8.5 and 9.5\,GHz respectively. Given the typical flux uncertainties and the probability that neither peak nor minimum flux are precisely sampled, this is consistent with a constant variation fraction with frequency.

%-------------------------------------------------------------------

\begin{figure*}
\centering
\includegraphics[width=1.85\columnwidth]{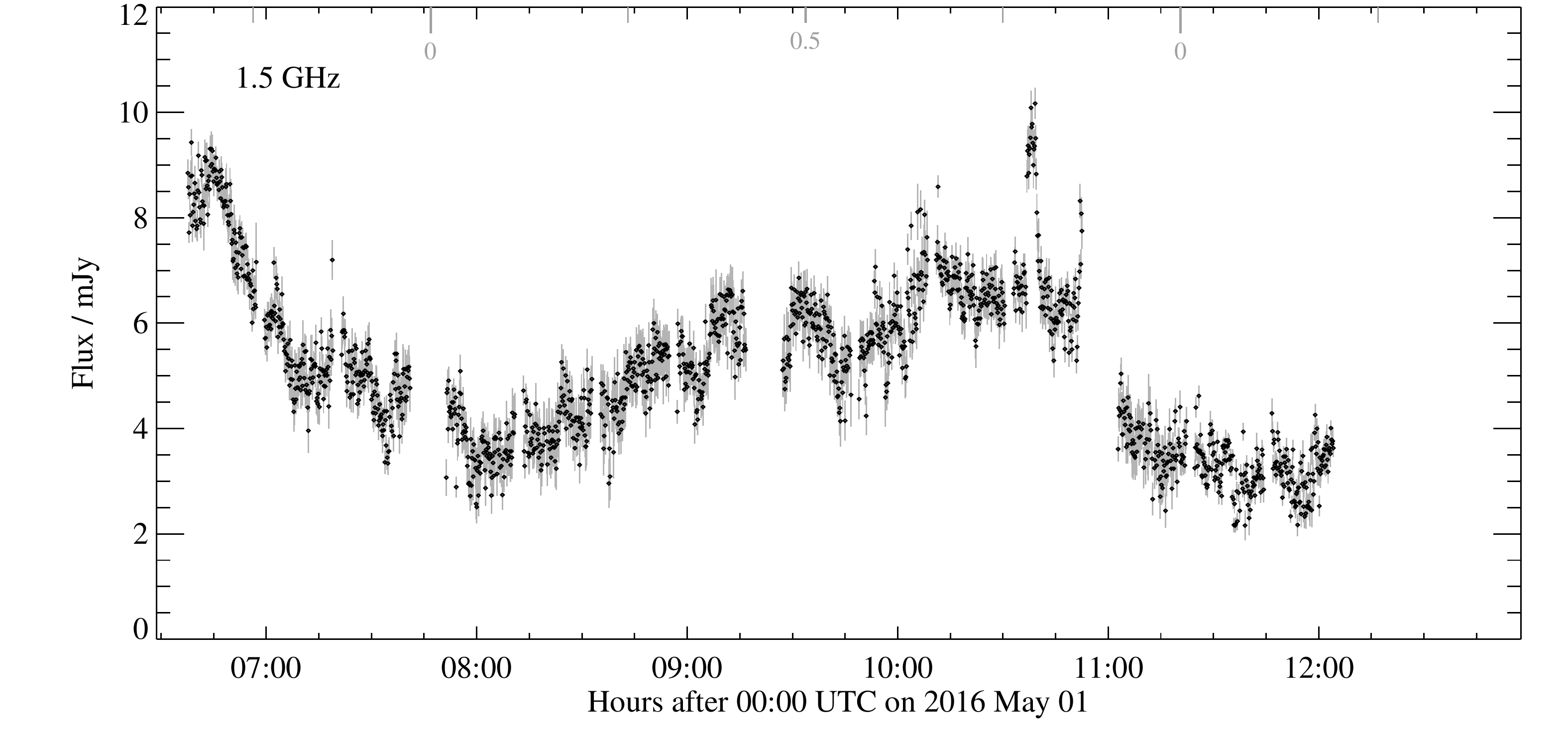}
\includegraphics[width=1.85\columnwidth]{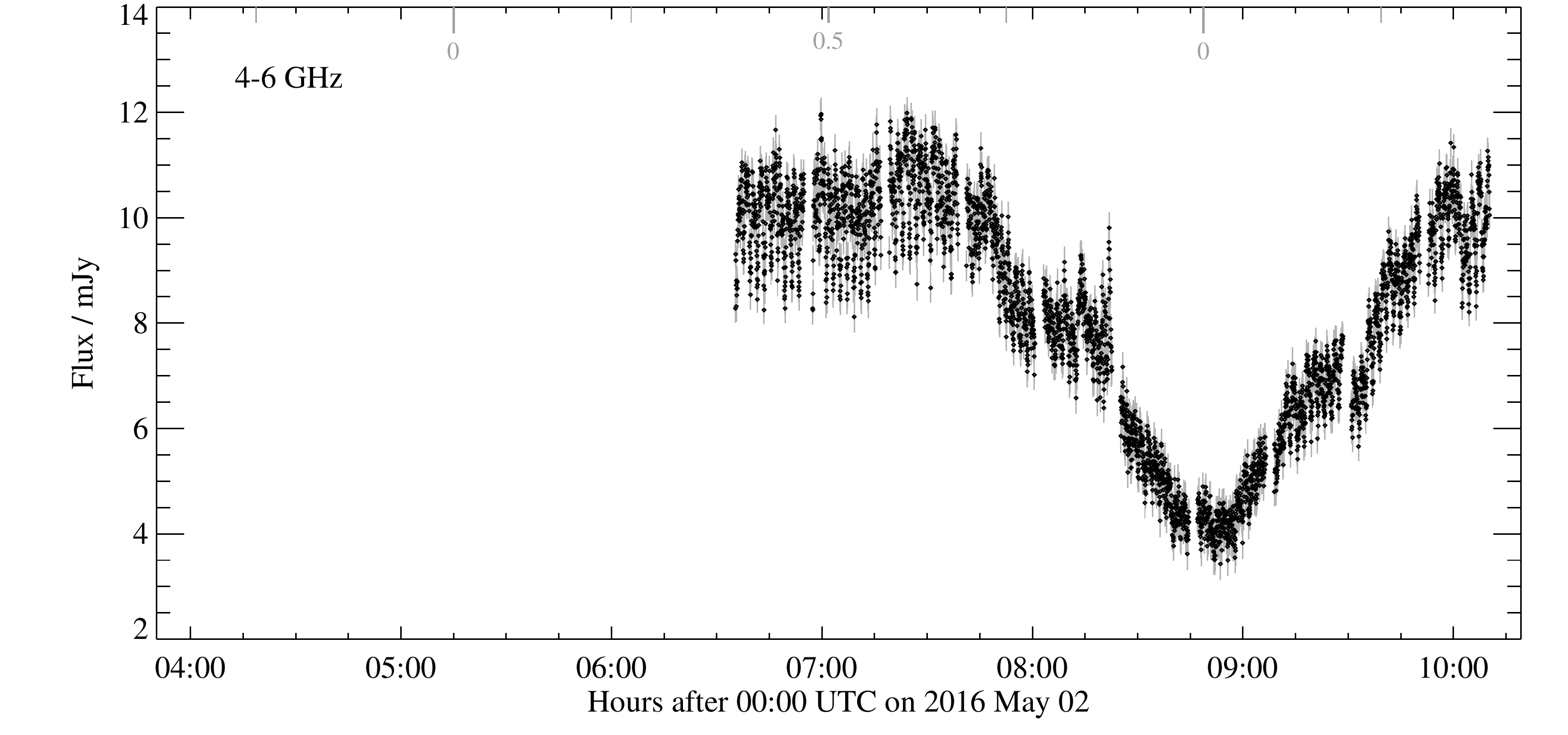}
\includegraphics[width=1.85\columnwidth]{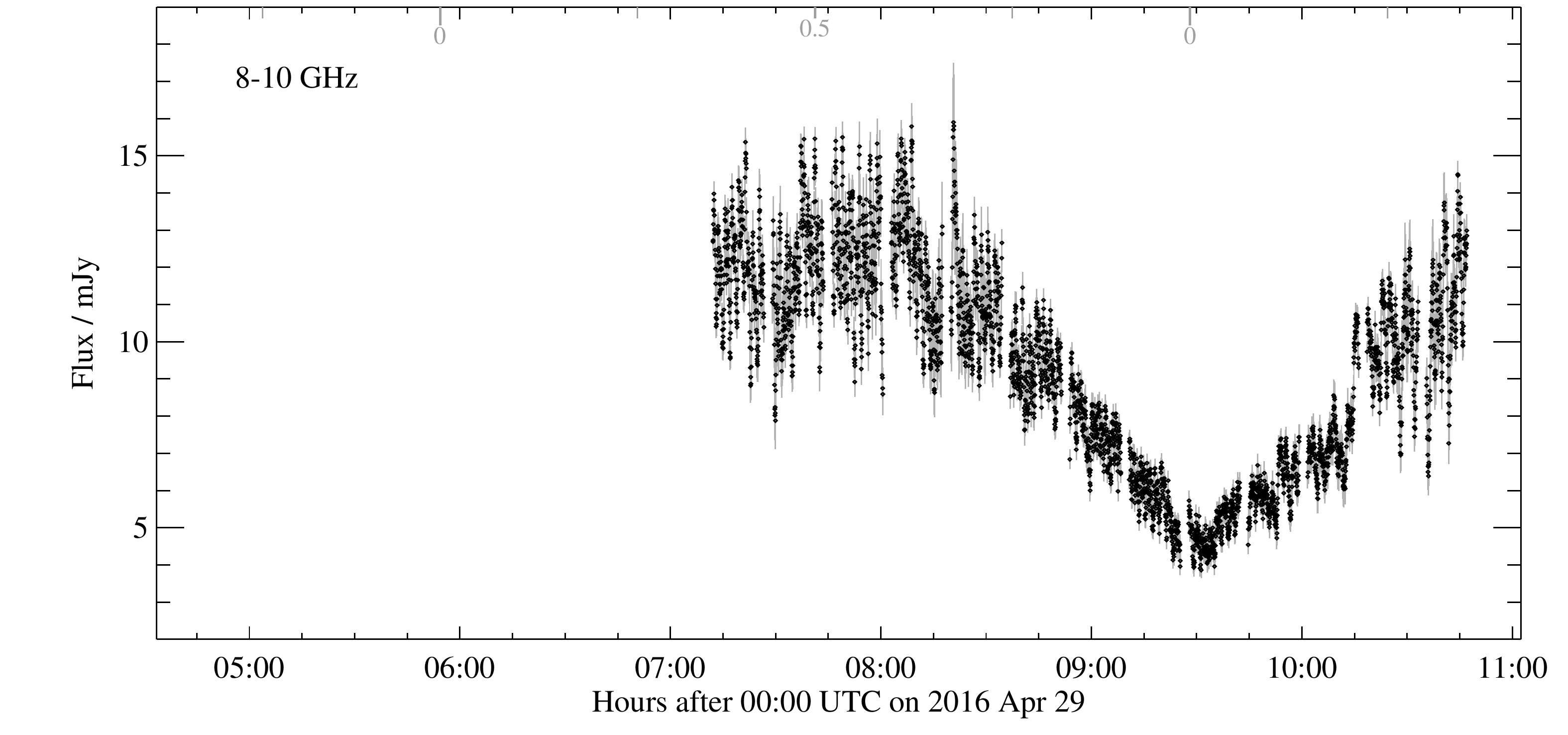}
\caption{Variation in the radio emission of \arsco\ on short timescales. Each data point is a flux extracted from an image constructed from 10\,s, 3\,s and 3\,s integrations in the 1.5, 5 and 9\,GHz bands respectively. Orbital phase is marked above the lightcurves.}
           \label{fig:short_lightcurves}%
 \end{figure*}

\begin{figure}
\centering
\includegraphics[width=0.75\columnwidth]{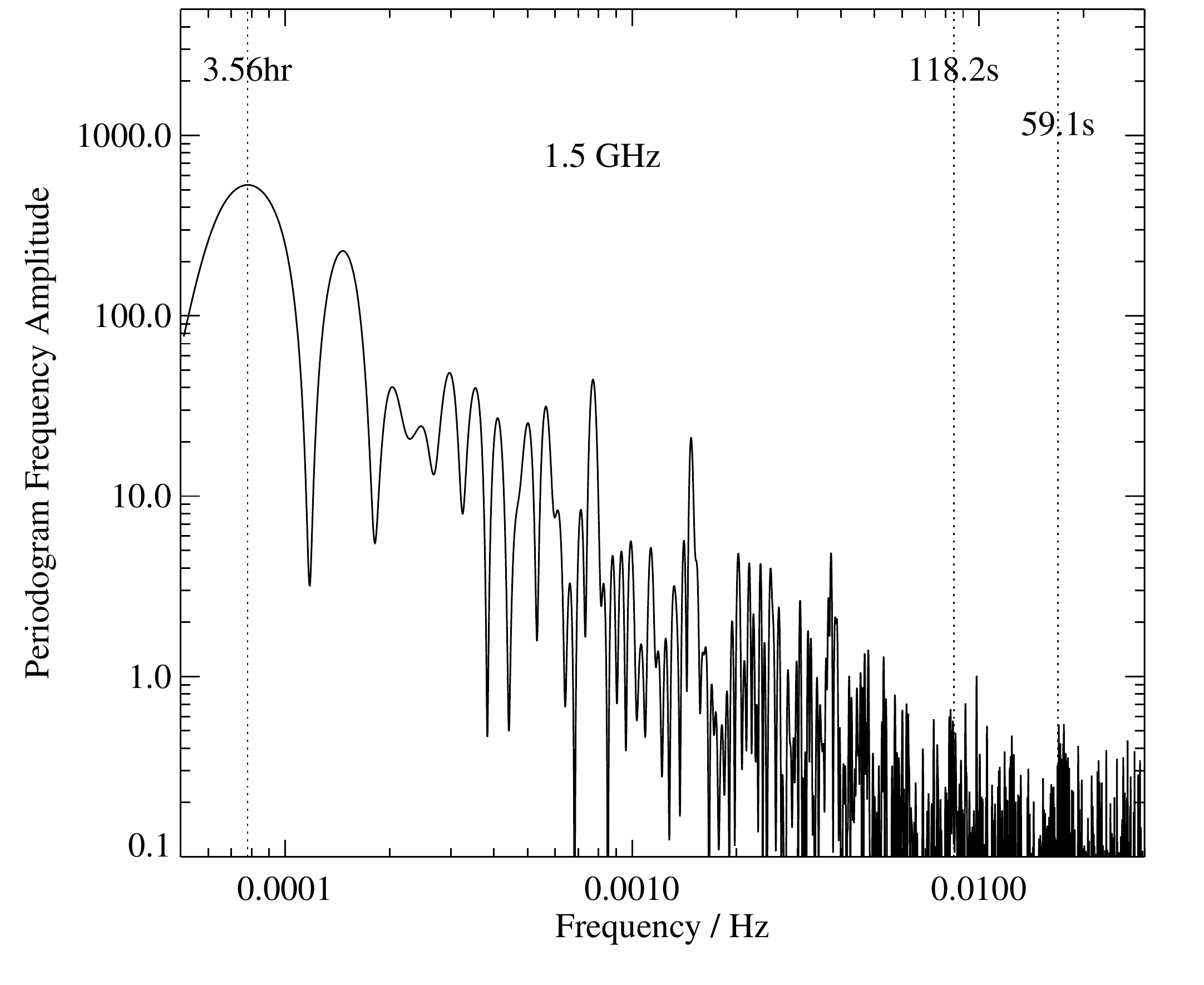}
\includegraphics[width=0.75\columnwidth]{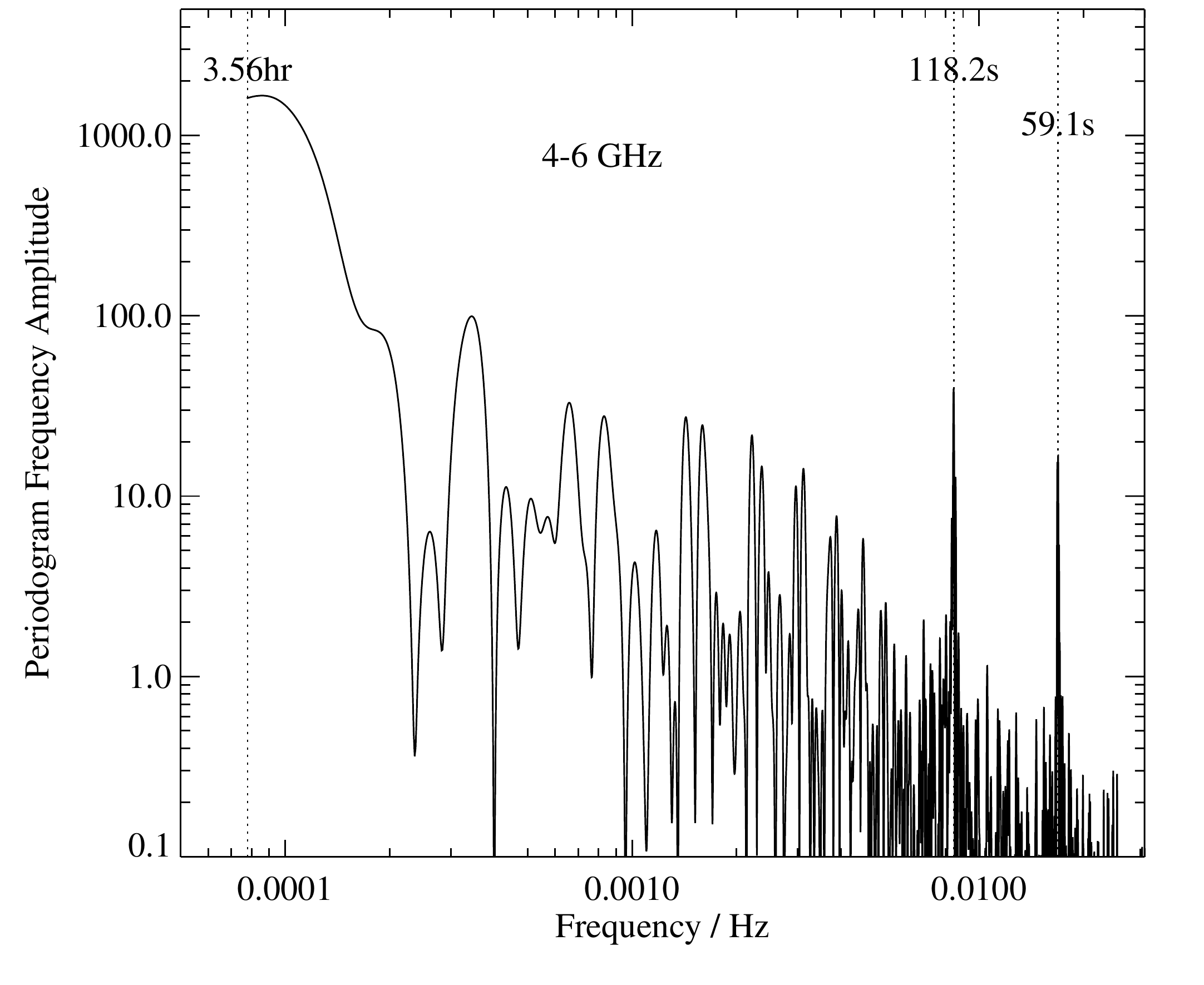}
\includegraphics[width=0.75\columnwidth]{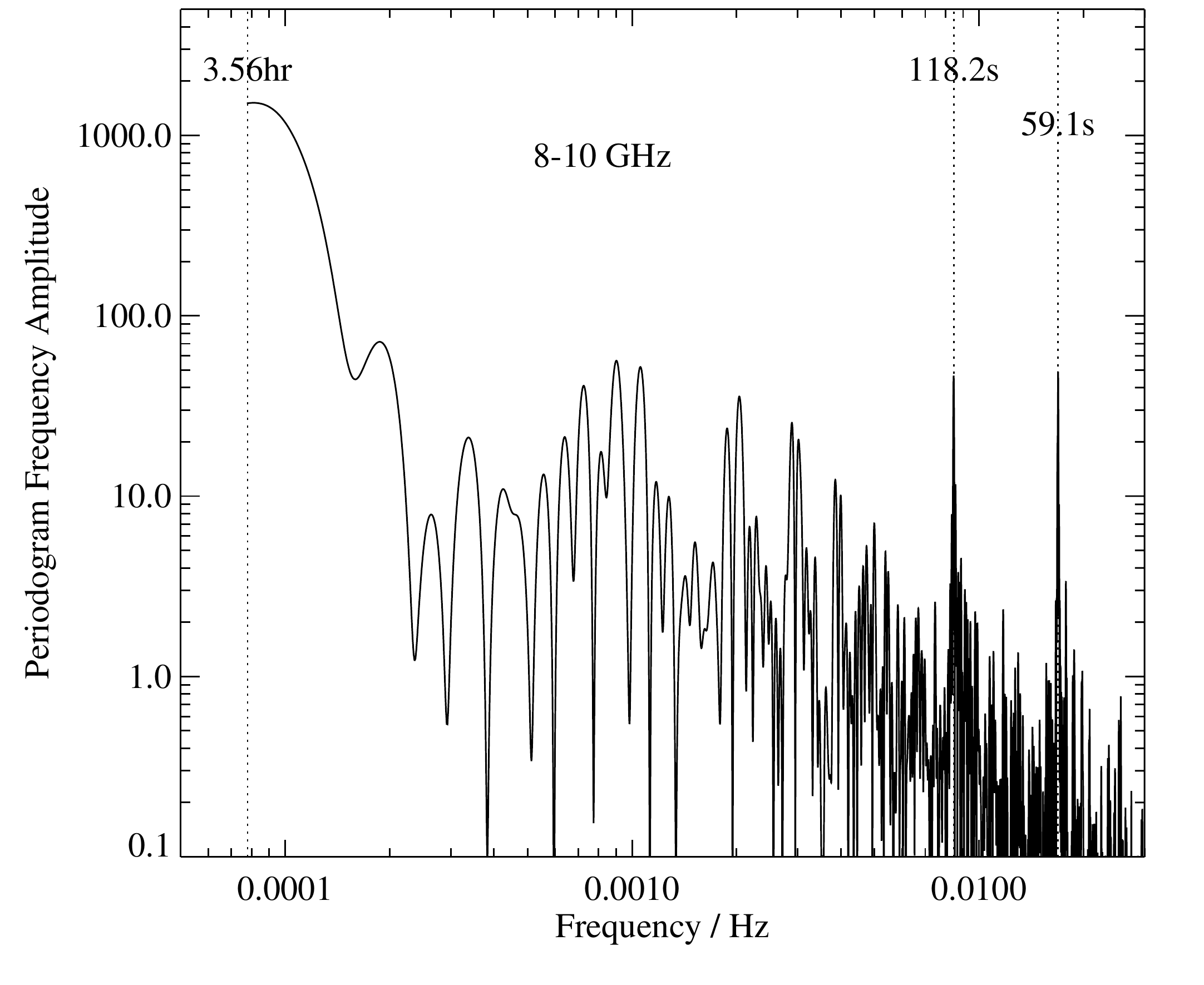}
\caption{Lomb-Scargle periodograms for the radio emission of \arsco\ as a function of frequency. The orbital period (3.56\,hr), system beat period (118.2\,s) and a half-beat period (59.1\,s) are indicated by vertical lines. 
}
           \label{fig:short_periodograms}%
 \end{figure}

\begin{figure*}
\centering
\includegraphics[width=1.2\columnwidth]{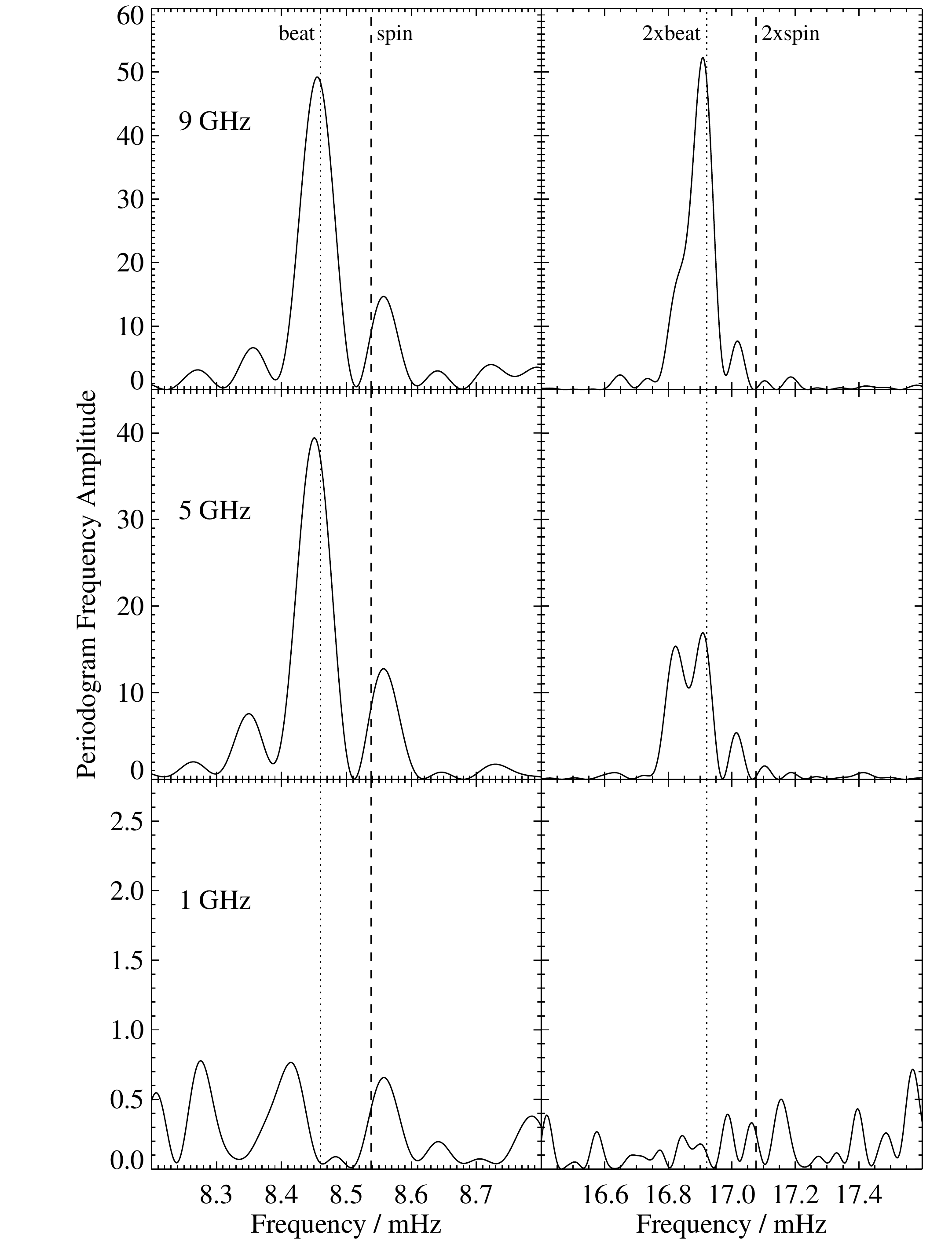}
\caption{Distinguishing beat from spin frequencies with the Lomb-Scargle periodograms for the radio emission of \arsco\ as a function of frequency.}
           \label{fig:periodogramzoom}%
 \end{figure*}

\begin{figure}
\centering
\includegraphics[width=0.75\columnwidth]{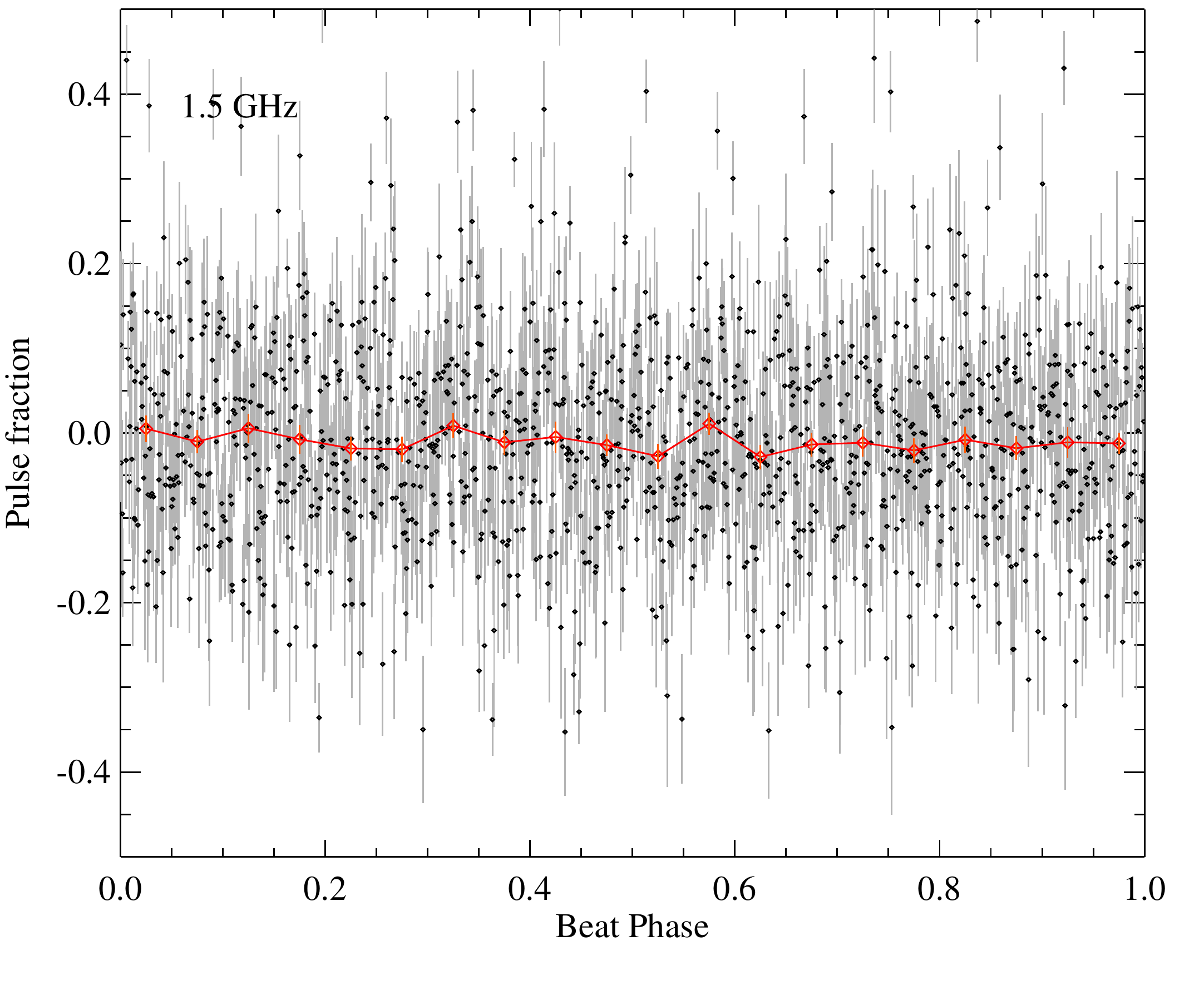}
\includegraphics[width=0.75\columnwidth]{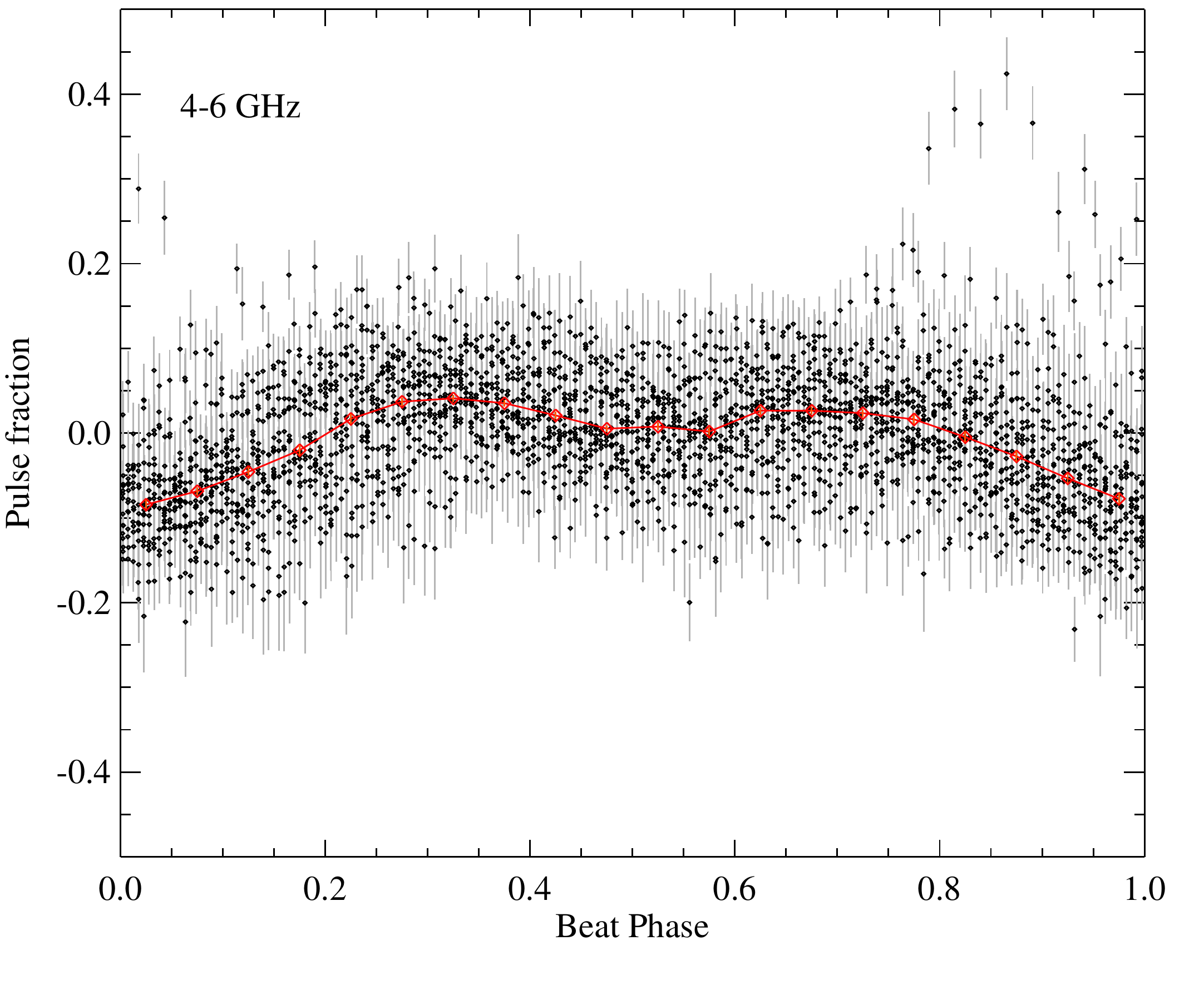}
\includegraphics[width=0.75\columnwidth]{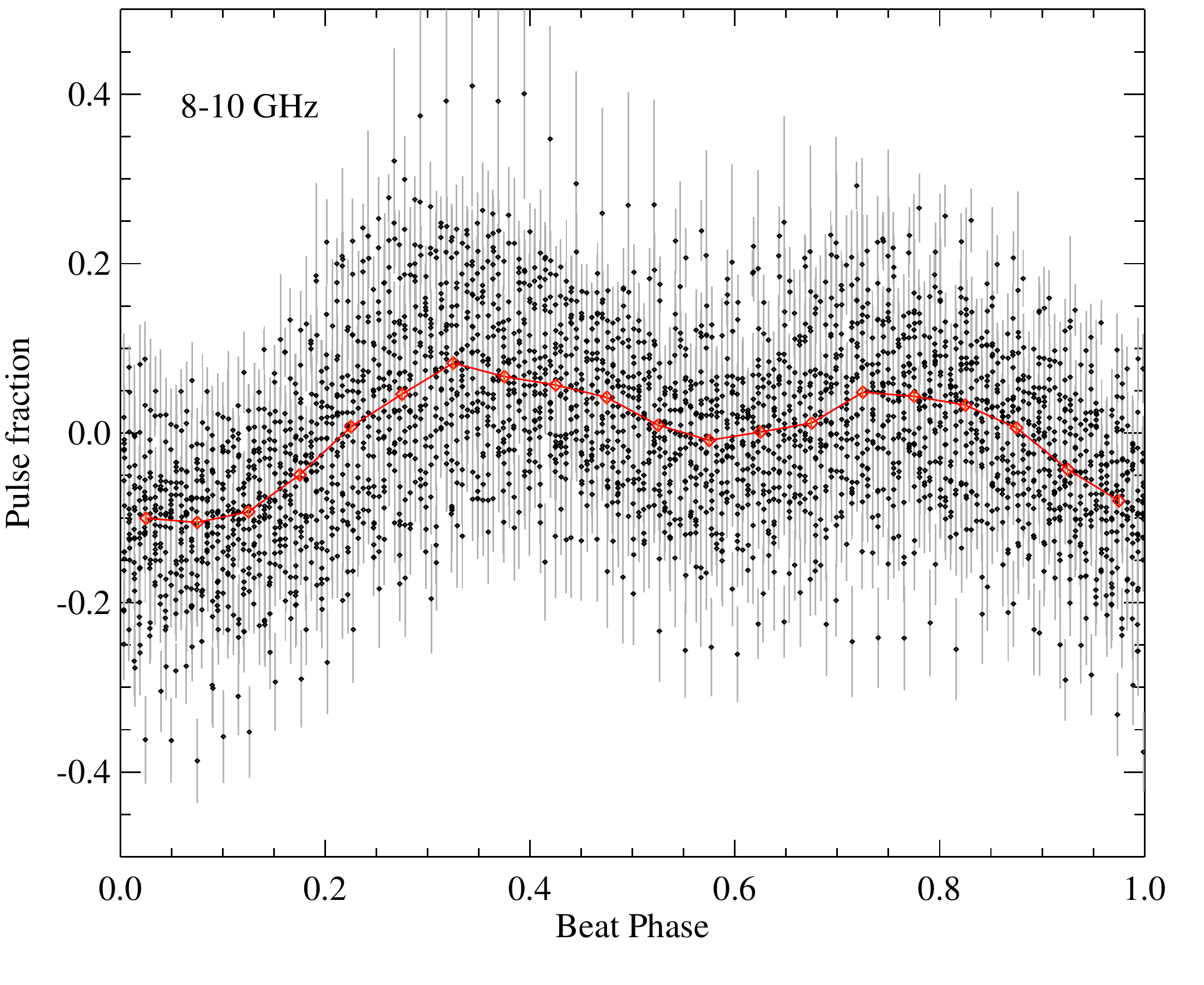}
\caption{Beat-folded lightcurves for the radio emission of \arsco\ as a function of frequency. In each band, the data has been phase-folded by the 118.199\,s beat period of the system, and each datum is normalised by a running mean over 450 seconds. Red points show the median in beat period phase bins of 0.05. Zero beat phase is defined at $T_0$(beat)=57510.28387390(BMJD), which corresponds to a minimum at 5\,GHz.}
           \label{fig:short_beatfolded}%
 \end{figure}

\begin{figure}
\centering
\includegraphics[width=0.75\columnwidth]{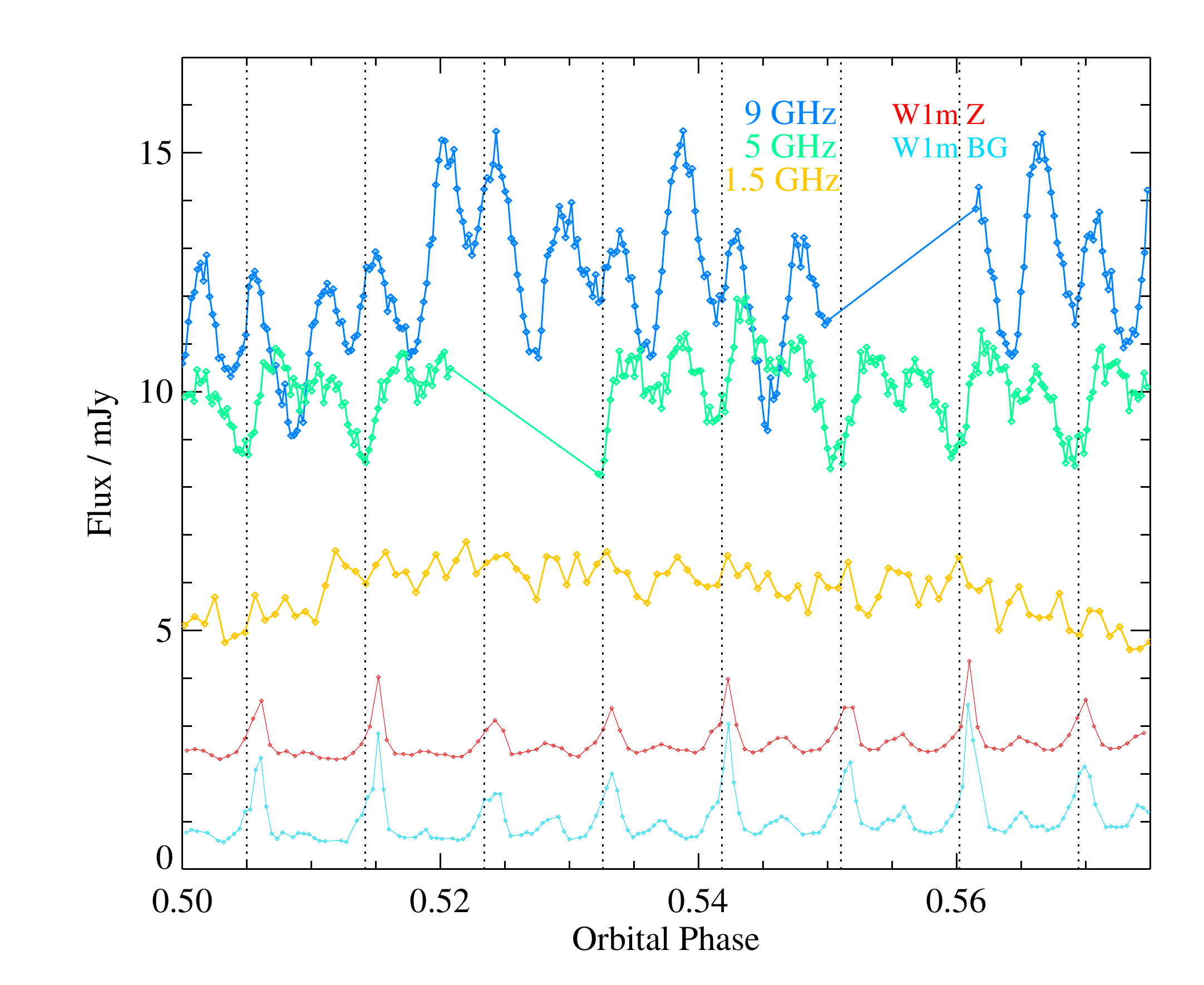}
\includegraphics[width=0.75\columnwidth]{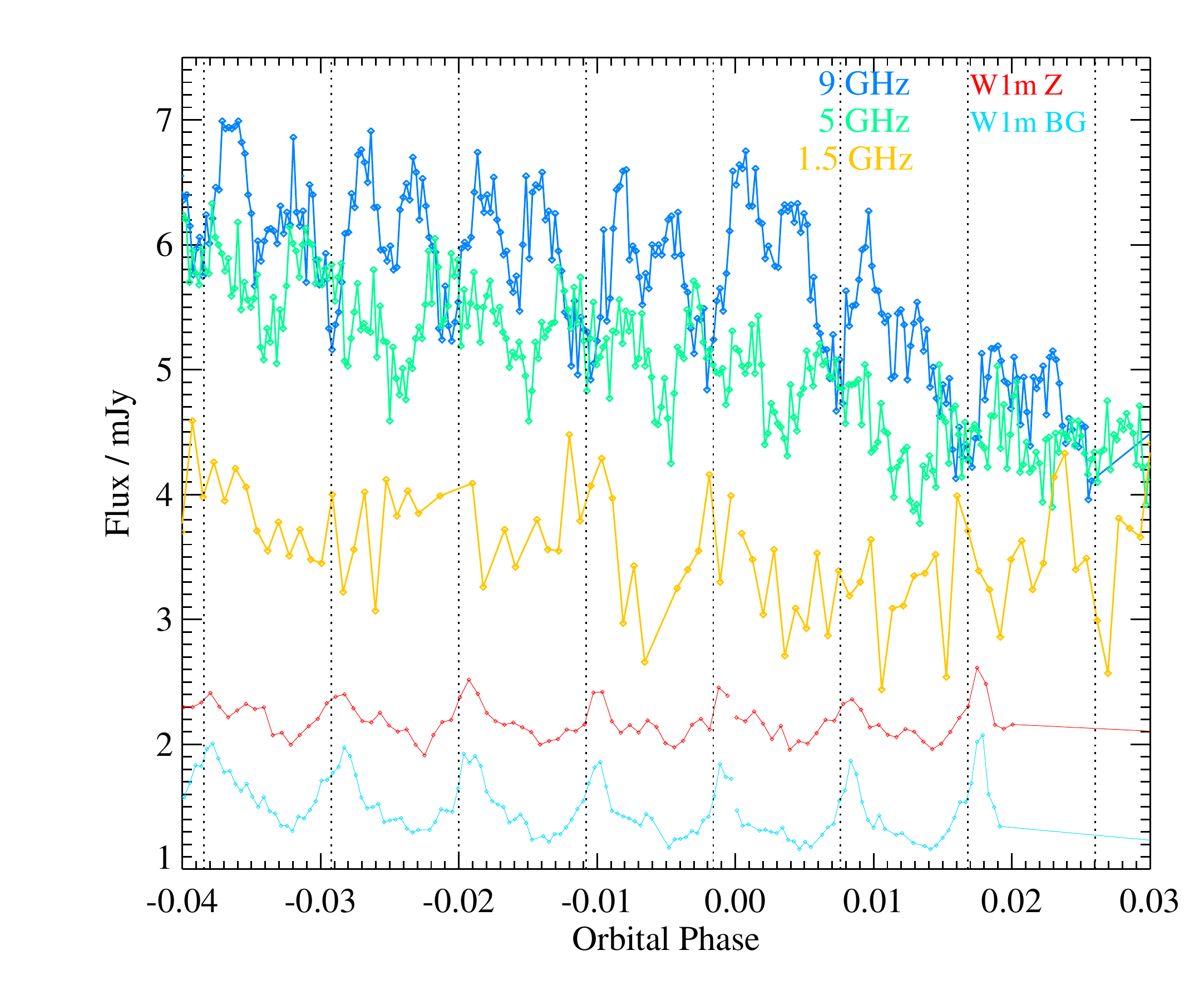}
\includegraphics[width=0.75\columnwidth]{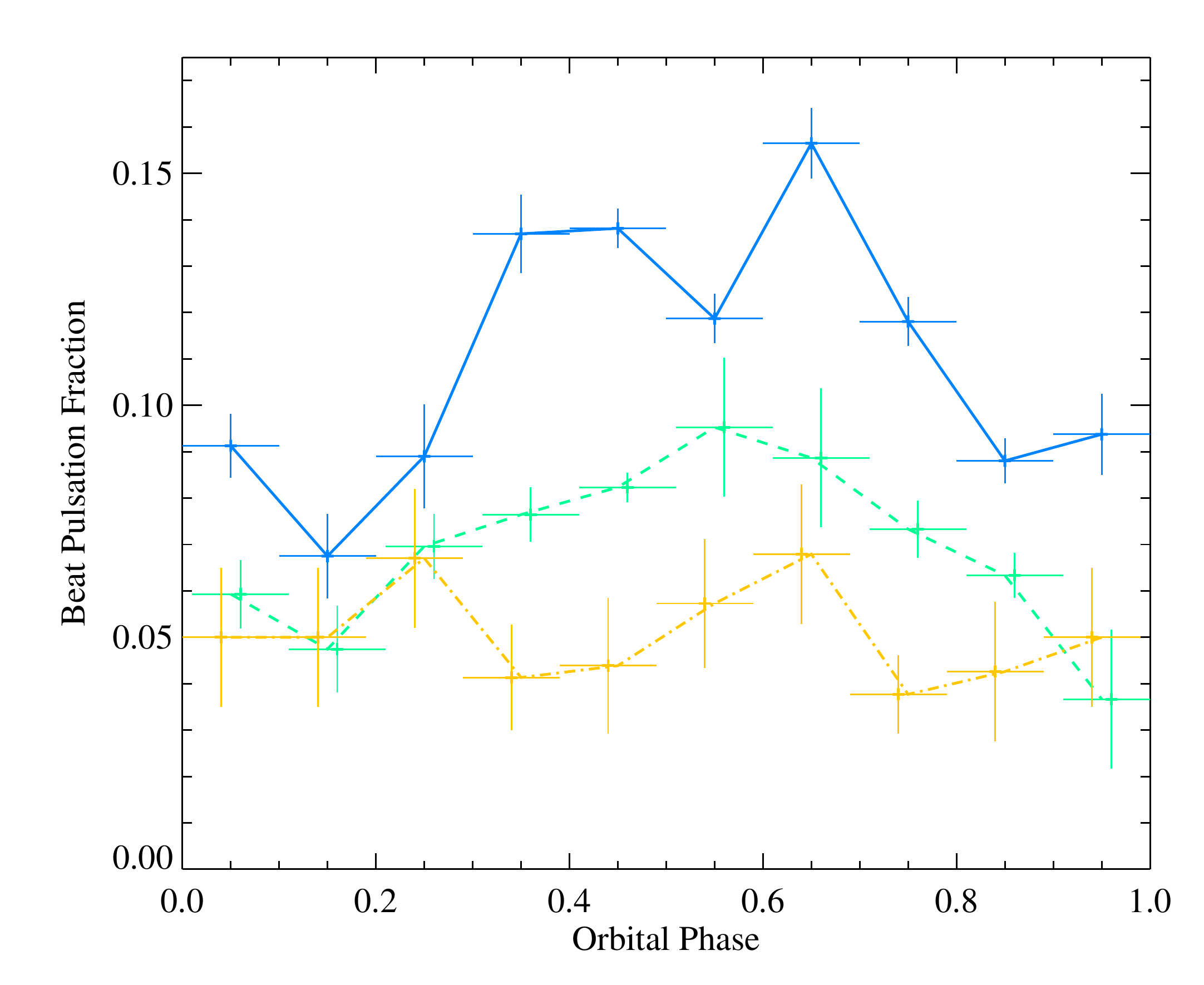}
\caption{The dependence of beat strength on orbital phase and frequency. We show that the beat pattern is captured in short integrations at the peak of the orbital lightcurve and at the orbital minimum. Vertical dotted lines mark intervals of 118.2\,s (the beat period), referenced to the 5\,GHz data as in Figure \ref{fig:short_beatfolded}. Optical data from W1m are shown with an arbitrary flux scaling for comparison. In the bottom panel, we indicate the half-amplitude of the phase-folded and binned beat lightcurve in each orbital phase bin, at 1.5, 5 and 9\,GHz.} % 
           \label{fig:beats}%
 \end{figure}

\begin{figure}
\centering
\includegraphics[width=0.75\columnwidth]{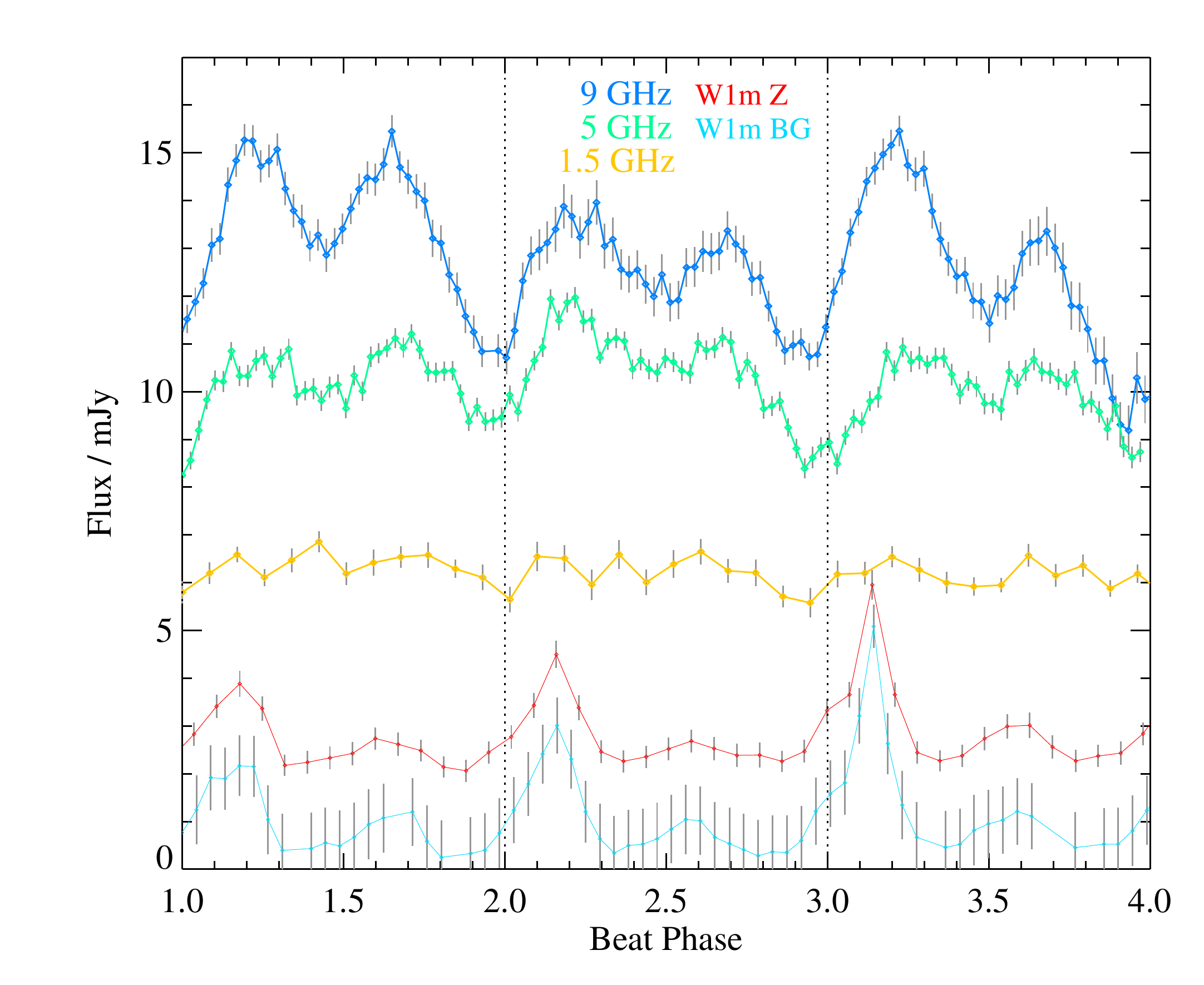}
\caption{The dependence of beat strength on beat phase and frequency. For each band, we show three consecutive beat cycles, extracted near the peak of the orbital light-curve and phased on the beat ephmeris. Only the W1m BG and Z bands are simultaneous, and these have an arbitrary scaling in flux, for comparison with the radio data. The significant beat-to-beat variations in the emission are apparent, as is a slight offset in beat phase between the optical and radio bands. Colour coding is as in figure \ref{fig:beats}.}            \label{fig:beatbeat}%
 \end{figure}

\section{Beat Period Properties}\label{sec:beat}

In order to explore the short time scale variation of \arsco\ it is necessary to image the radio data, and measure the source flux, at very short intervals. Fortunately, the brightness of the source together with the very good $uv$-plane support for snapshot observations at the VLA allow this. We use a custom CASA python script to generate a {\sc CLEAN}ed image and to run the {\sc IMFIT} task on the visibility data for integration intervals of 3\,s in the 5 and 9\,GHz bands and 10\,s in the 1.5\,GHz band (where the source is relatively faint and the beam is large). In each case, the full bandwidth available (2\,GHz, 2\,GHz, 1\,GHz respectively) is used. \arsco\ is well detected in all these images and  {\sc IMFIT} reports that the source is a point source (as expected).

As before we fit \arsco\ alone in the 5 and 9\,GHz bands, and the target and neighbouring source simultaneously in the L band. In a handful of intervals, the fit failed to deconvolve the two sources, but in the majority the flux was successfully measured. The detailed lightcurve of \arsco\ in each frequency band is shown in Figure \ref{fig:short_lightcurves}, and demonstrates the extraordinary variability of \arsco\  on short timescales. This includes aperiodic flaring, particularly apparent in the 1.5\,GHz band, in addition to the known pulsations of the system. While the flares seen at 1.5\,GHz have not been reported in the optical, they show timescales of just a few minutes, and it is likely that a similar (20\% of total flux) flare would be masked by the strong beat/spin modulation in the optical.

In Figure \ref{fig:short_periodograms} we calculate a Lomb-Scargle periodogram \citep{1976Ap&SS..39..447L,1982ApJ...263..835S,1989ApJ...338..277P} of each dataset, in order to determine the dominant frequency components in the variability. Unsurprisingly, there is a strong  component at the orbital frequency in each band. There is also a signal at high frequencies. In Figure \ref{fig:periodogramzoom}, we examine this region, and twice its frequency, in more detail.  At both 5 and 9\,GHz we identify significant power at a frequency consistent with the beat period ($P_B=118.2$\,s) between the binary orbit and the white dwarf spin ($P_S=117.0$\,s, Marsh et al 2016). There is no clear evidence for emission on the spin period suggesting that the dominant power source for the radio emission lies in an interaction between the white dwarf and its red dwarf companion. The beat signal appears to decrease in strength with frequency. In the 1.5\,GHz band, this beat signal disappears, and the only significant peak in the power spectrum corresponds to the orbital period, while the secondary peaks appear to be related to the length of individual scans with the VLA. To verify this, we have performed a test, subtracting a simple sinusoidal orbital modulation from the 1.5\,GHz data. The apparent periodogram features at about 10 and 20 minutes period are heavily suppressed in a periodogram of the residuals, confirming that they are associated with the window function of the data.

Phase-folding on the beat period, we are able to recover the beat-pulse lightcurve at each frequency. A reference zero beat phase is defined at $T_0$(beat)=57510.28387390 (BMJD), which corresponds to a minimum at 5\,GHz. This is shown in Figure  \ref{fig:short_beatfolded}, where every datapoint has been normalised relative to the underlying orbital flux variation, determined by calculating a running mean across 450 seconds in each band. The plotted pulsation fraction is therefore defined by $f_\mathrm{obs,i}/f_\mathrm{mean,\Delta i}$. 

In the 5\,GHz and 9\,GHz bands, this pulsation fraction varies throughout the orbital period, reaching a maximum of about $\pm20$\% with a 5\% uncertainty. Taking the median value at each phase, the typical pulsation varies from +4 to -8\% around the mean flux in the beat cycle at 5\,GHz and from +8 to -10\% at 9\,GHz. The beat oscillation is itself double peaked on the beat period, but asymmetric, with the two peaks of unequal strength. This behaviour was also seen in the near-infrared, optical and ultraviolet (see Marsh et al 2016, Figure 2).

As Figures \ref{fig:beats} and \ref{fig:beatbeat} demonstrate, the same asymmetric double pulse within a beat period is identifiable in individual beat cycles at the peak of the orbital lightcurve, and is clearly detected at 5 and 9\,GHz, with no apparent lag between them at a given beat phase. The ratio of the two pulses varies significantly from pulse to pulse, as exemplified by the example sequences of three consecutive beats at each frequency in figure \ref{fig:beatbeat}. The two beats in each period are typically closer in strength at 5\,GHz than at 9\,GHz, and not nearly so asymmetric as seen in the optical. There also appears to be a slight phase lag between the peaks at optical frequencies and those in the radio.

The pulsations fade with decreasing frequency, and cannot be clearly seen in the 1.5\,GHz data, even at peak orbital flux. At the minimum of the orbital lightcurve (orbital phase = 0), it becomes harder to identify pulsations at all frequencies, primarily due to reduced signal to noise in the individual integrations. By subdividing the data first on orbital phase, and then on beat phase, and calculating the median in 0.05 beat phase bins (as in Figure \ref{fig:short_beatfolded}), it is possible to look at how the pulsation strength varies with orbital phase, and we show this in the bottom panel of Figure \ref{fig:beats}. Here we use the half-difference between the maximum and minimum in the orbit- and beat-folded, median-averaged lightcurve as a proxy for pulsation strength.
The strongest pulsations, varying by 15\% relative to the median flux in a 0.1 orbital phase (21.4 minute) bin, are  seen in the 9\,GHz data at an orbital phase around 0.5-0.6. The 5\,GHz data shows a similar strengthening of the pulse fraction at an orbital phase of $\sim$0.5. The 1.5\,GHz data do not show any evidence for periodic variation in the folded lightcurves, and the typical 10\% difference between maximum and minimum in these lightcurves reflects the signal to noise of the data.

%-------------------------------------------------------------------

\section{Polarization}\label{sec:pol}

In addition to total flux measurements (the Stokes $I$ parameter), the VLA data contains full polarization information. Observations were taken at each frequency of the low-polarization leakage calibrator J1407+2827 as part of the relevant scheduling block. These were used to determine the instrumental polarization, while the known polarization properties of our primary flux calibrator 3C286 were used to determine the cross-hand delays and the R-L polarization angle. Standard CASA tasks {\sc gaincal} and {\sc polcal} were used to calculate these, and they were applied together with the basic gain, delay and flux amplitude calibration tables generated by the VLA pipeline.

\subsection{Polarization with Orbital Phase}

\subsubsection{9\,GHz band}
We first explore polarization data at 9\,GHz (X band), where the total flux and beat signal are strongest. In Figure \ref{fig:polorbit9} we illustrate the polarization with orbital phase. Data were initially imaged by (15 minute) scan in all four Stokes parameters. Once the variation was identified in the Stokes $V$ parameter, each scan was divided into two six minute periods (where an even number of minutes was chosen to ensure consistent averaging over the beat period). Note, due to the necessity of phase calibration and the non-integer number of beat periods per scan, there is an unobserved interval between each pair of integrations. As before, the CASA {\sc imfit} function was used to determine the flux of \arsco\ in each image, and where the fit failed due to the lack of any visible source, a flux of zero is assigned in that Stokes parameter. Typical uncertainties are of order 0.2\% in the polarization, dominated by the RMS uncertainty in the polarized images, rather than the total Stokes $I$ flux.

The resultant polarization light curve shows a strong dependence on orbital phase.  The linear polarization is consistently low, with either no detectable source in the Stokes parameter imaging or a point source detection at $<2\,\sigma$ over the bulk of the orbital period.  However at orbital phases between -0.05 (0.95) and 0.2, the linear polarization is consistently above zero (albeit at low significance). The dominant component switches between $Q$ and $U$, suggesting a rapid rotation of the linear polarization (on a timescale of minutes). At its maximum, the source has a total linear polarization (in a six minute interval) of $1.0\pm0.3$\%.

By contrast both the strength and variation of circular polarization (the Stokes $V$ parameter) in \arsco\ are more extreme. For the bulk of the orbital period, the source is consistent with exhibiting negative (anticlockwise) circular polarization. This polarization becomes highly significant at an orbital phase of $\approx$0.95 with $V=-5.7\pm0.2$\%. It exhibits a secondary (lower significance) negative peak with $V=-0.23\pm0.06$\% at an orbital phase of $\approx$0.65.

\subsubsection{5\,GHz band}
This behaviour is mirrored in the polarization at 5\,GHz (C band), where we integrate over individual 10 minute scans. As before, the source is consistent with a very low linear polarization, peaking short of 1\% at an orbital phase close to zero. By contrast, the circular polarization shows a very strong orbital modulation, peaking at $V=-7.7\pm0.3$\%, at an orbital phase of 0.9, somewhat earlier than that seen at 9\,GHz. The secondary peak, at an orbital phase of 0.65, is also reproduced and is well-detected, reaching $V=-1.3\pm0.2$\%.

\subsubsection{1.5\,GHz band}

The orbital behaviour of polarization at 1.5\,GHz (L band) is far more extreme than that seen at higher frequencies. Again we integrate over 10 minute scans, and do not attempt to subdivide these due to the relatively low total flux signal-to-noise in this band. The lightcurve at 1.5\,GHz spans more than a full orbital period, giving more indication of the orbit-to-orbit variation. The linear polarization is in the range 1-3\% for the majority of the orbit, reaching a minimum around an orbital phase of 0.45 and showing a maximum of $\approx3$\%, close to an orbital phase of 1 (slightly after this in the first orbit, slightly before in the second, but consistent given the random uncertainties on each data point).

The circular polarization shows a similar excursion to large negative values to that seen at higher frequencies, but reaching $V=-27\pm1$\% in the first orbit and $V=-22\pm1$\% in the second orbit. This negative peak appeared earlier at 5 than 9\,GHz. It appears earlier still at 1.5\,GHz and is broader, with a significant negative excursion in polarization between phases of 0.6 and 0.8. This feature also shows internal structure, splitting into two overlapping negative peaks, rather than being a single narrow peak as seen at higher frequencies. The secondary peak seen at 5 and 9\,GHz is also apparent at 1.5\,GHz, again shifted earlier in the orbit.

\begin{figure}
\centering
\includegraphics[width=0.8\columnwidth]{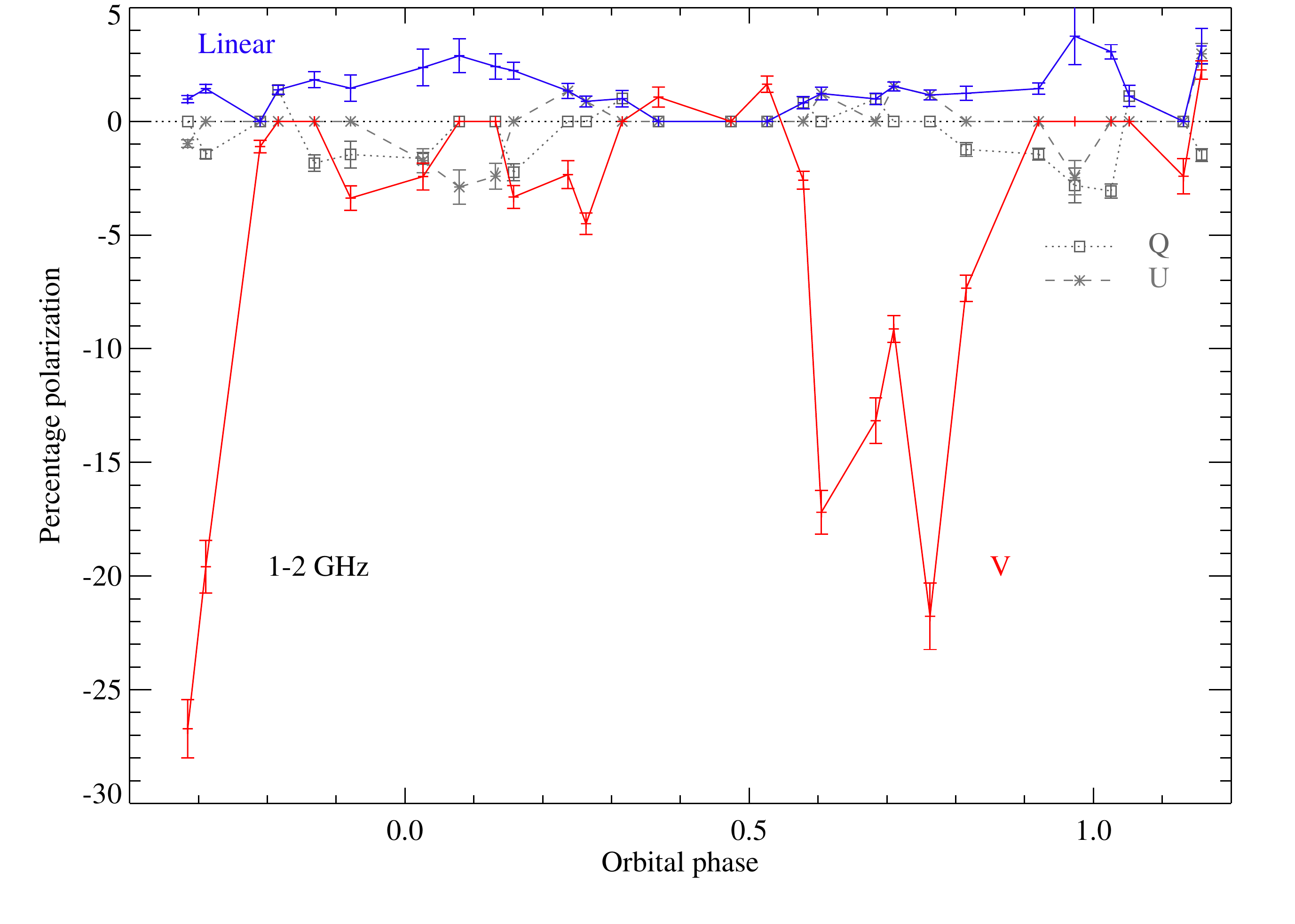}
\includegraphics[width=0.75\columnwidth]{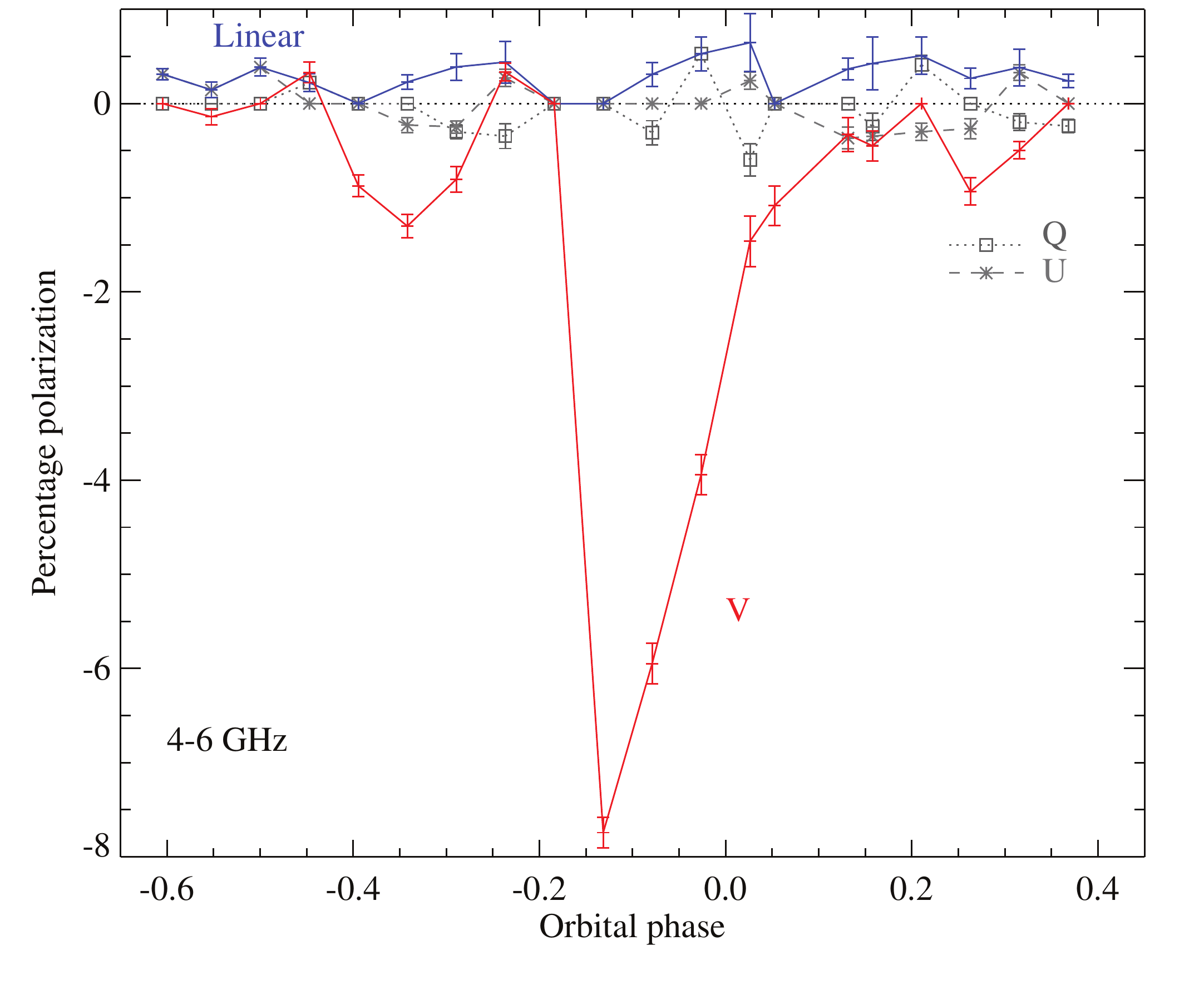}
\includegraphics[width=0.75\columnwidth]{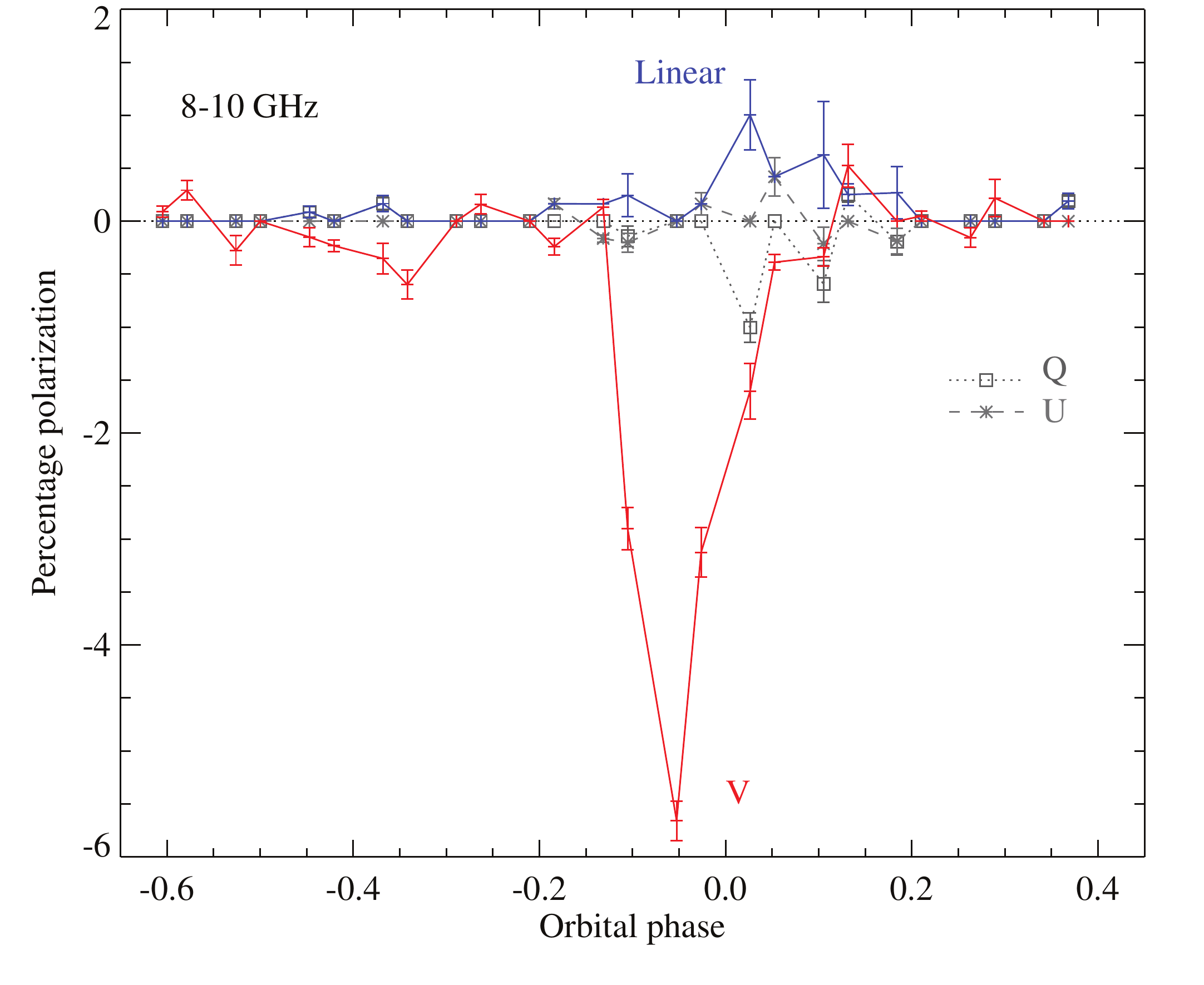}
\caption{The dependence of polarization on orbital phase at 1.5\,GHz (upper panel), 5\,GHz (middle panel) and 9\,GHz (lower panel). Data were imaged in all four Stokes parameters in 10, 10 and 6 minute time bins respectively (i.e. averaging over beat period behaviour). We show the circular polarization percentage (Stokes V/I) and the total linear polarization percentage ($\sqrt{(Q^2+U^2)/I^2}$), as well as the Stokes Q and U terms. An identically zero polarization indicates that {\sc imfit} was unable to identify a source at the location of \arsco\ in that Stokes parameter.}
           \label{fig:polorbit9}%
 \end{figure}

\begin{figure}
\centering
\includegraphics[width=0.75\columnwidth]{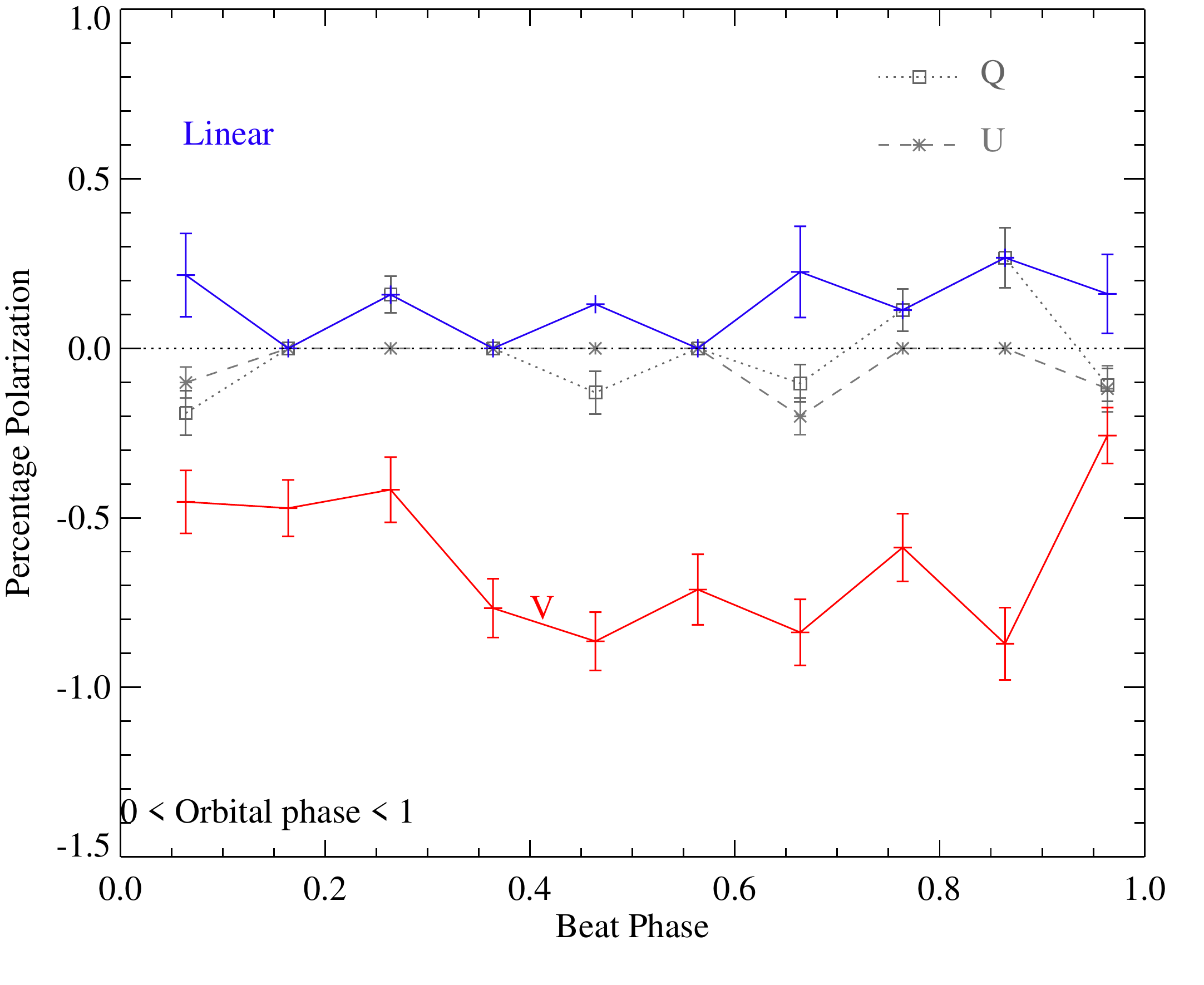}
\includegraphics[width=0.75\columnwidth]{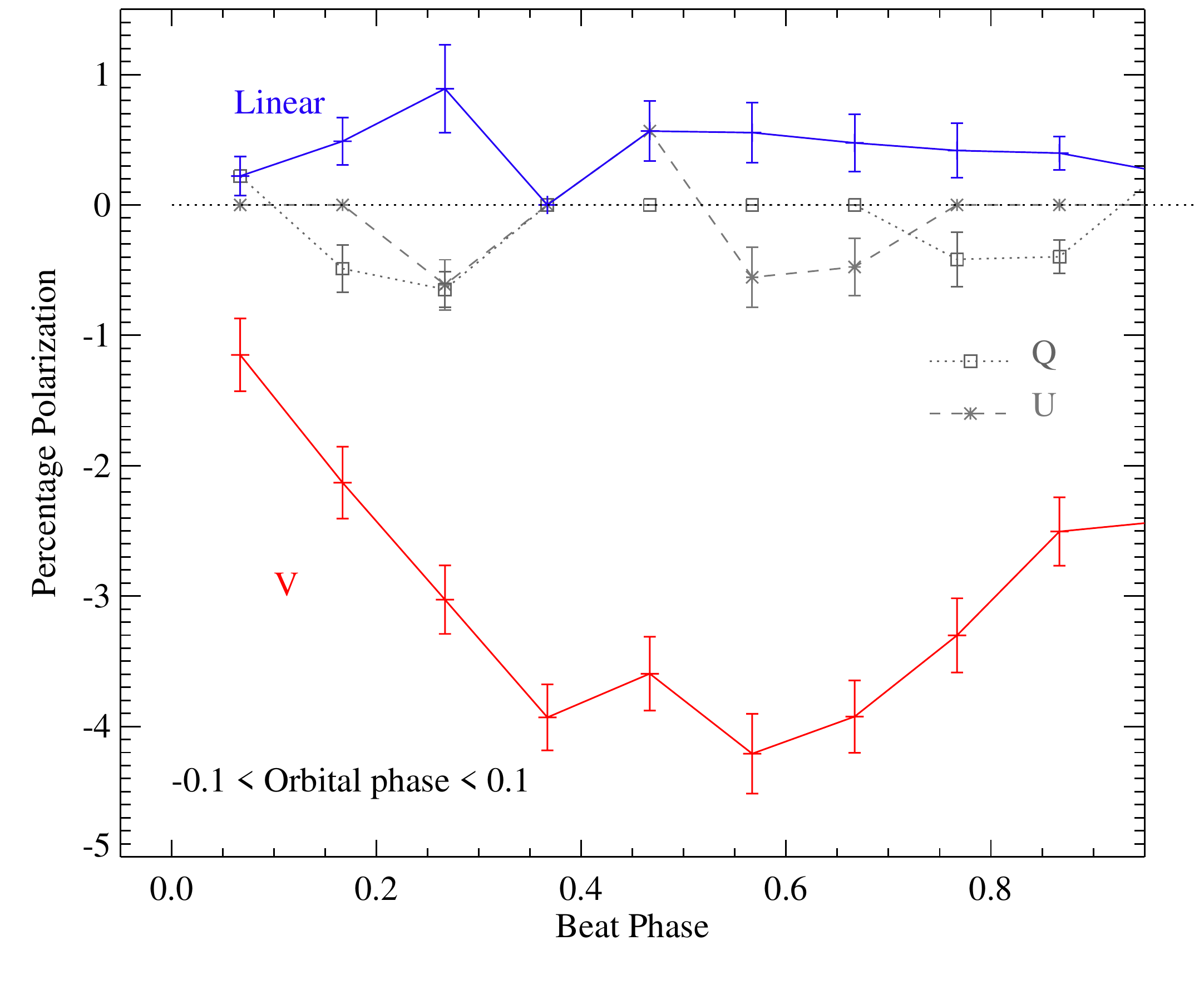}
\includegraphics[width=0.75\columnwidth]{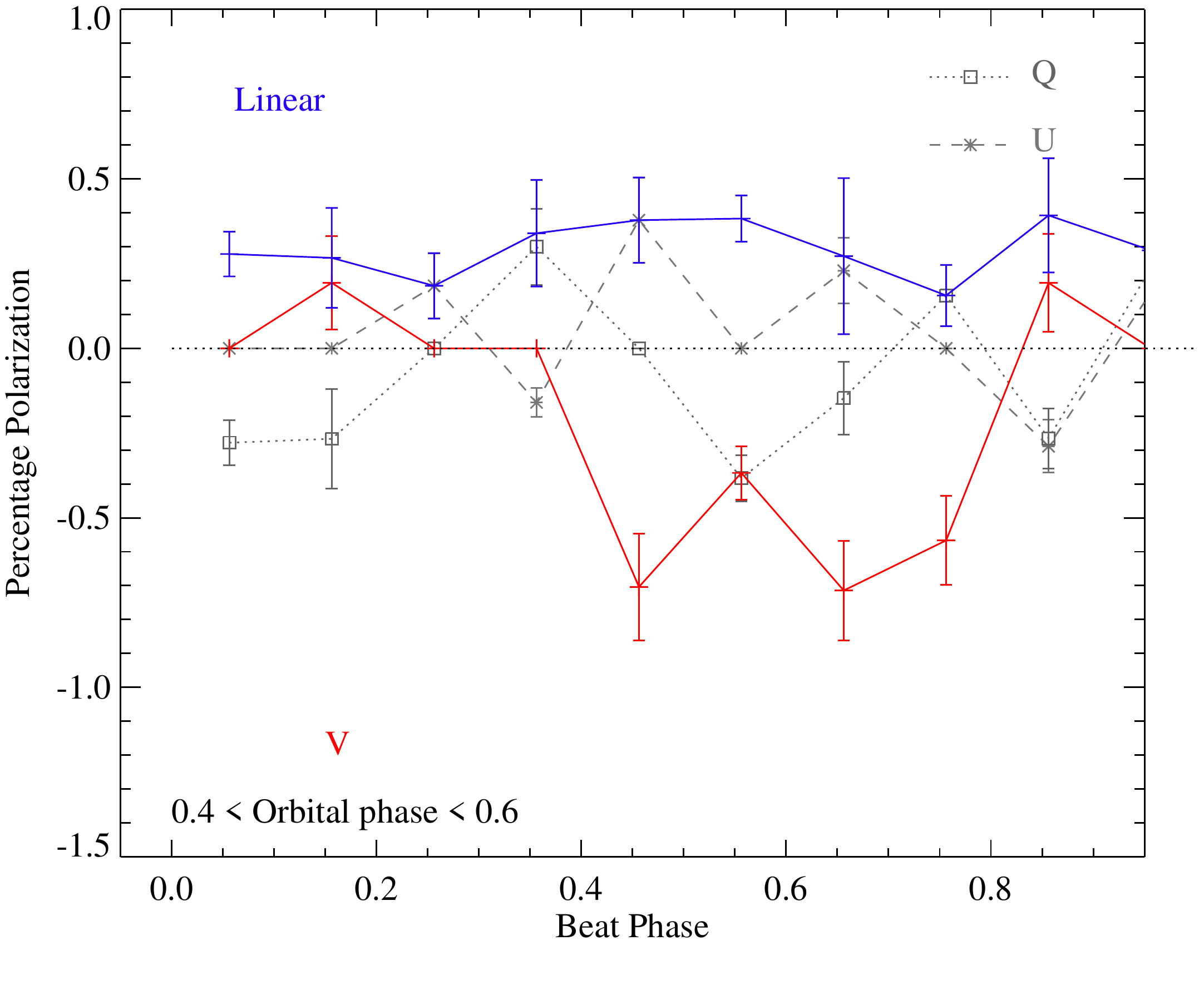}
\caption{The dependence of polarization on orbital phase and beat phase at 9\,GHz. Visibility data were folded on the beat period, and each bin of 0.1 in beat phase seperately imaged in $IQUV$. Beat phase zero is defined as in figure \ref{fig:short_beatfolded}. This procedure was repeated for two subsets at orbital phases $\approx0$ and 1. An identically zero polarization indicates that {\sc imfit} was unable to identify a source at the location of \arsco\ in that Stokes parameter.}
           \label{fig:pol_beats9}%
\end{figure}

\begin{figure}
\centering
\includegraphics[width=0.75\columnwidth]{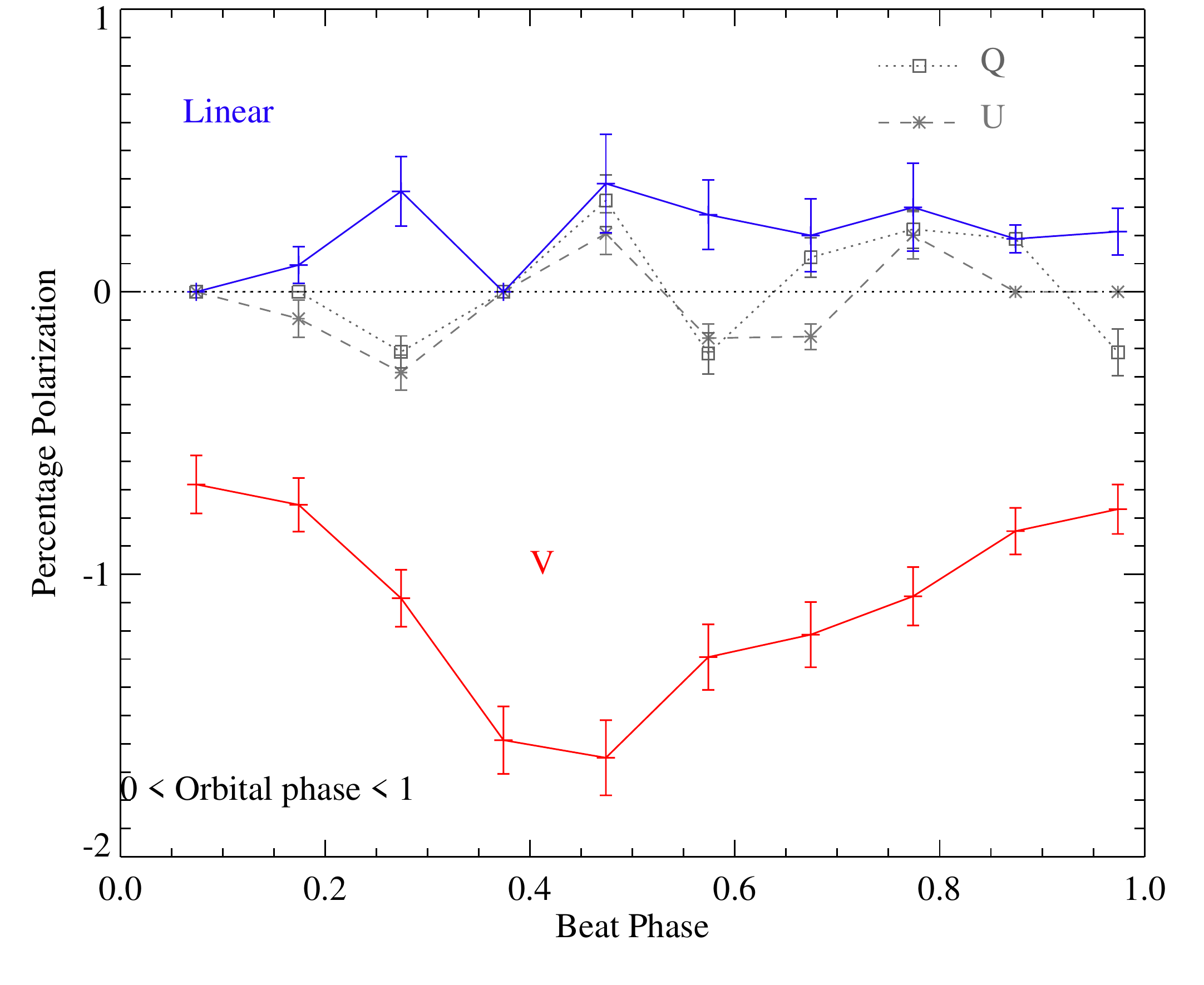}
\includegraphics[width=0.75\columnwidth]{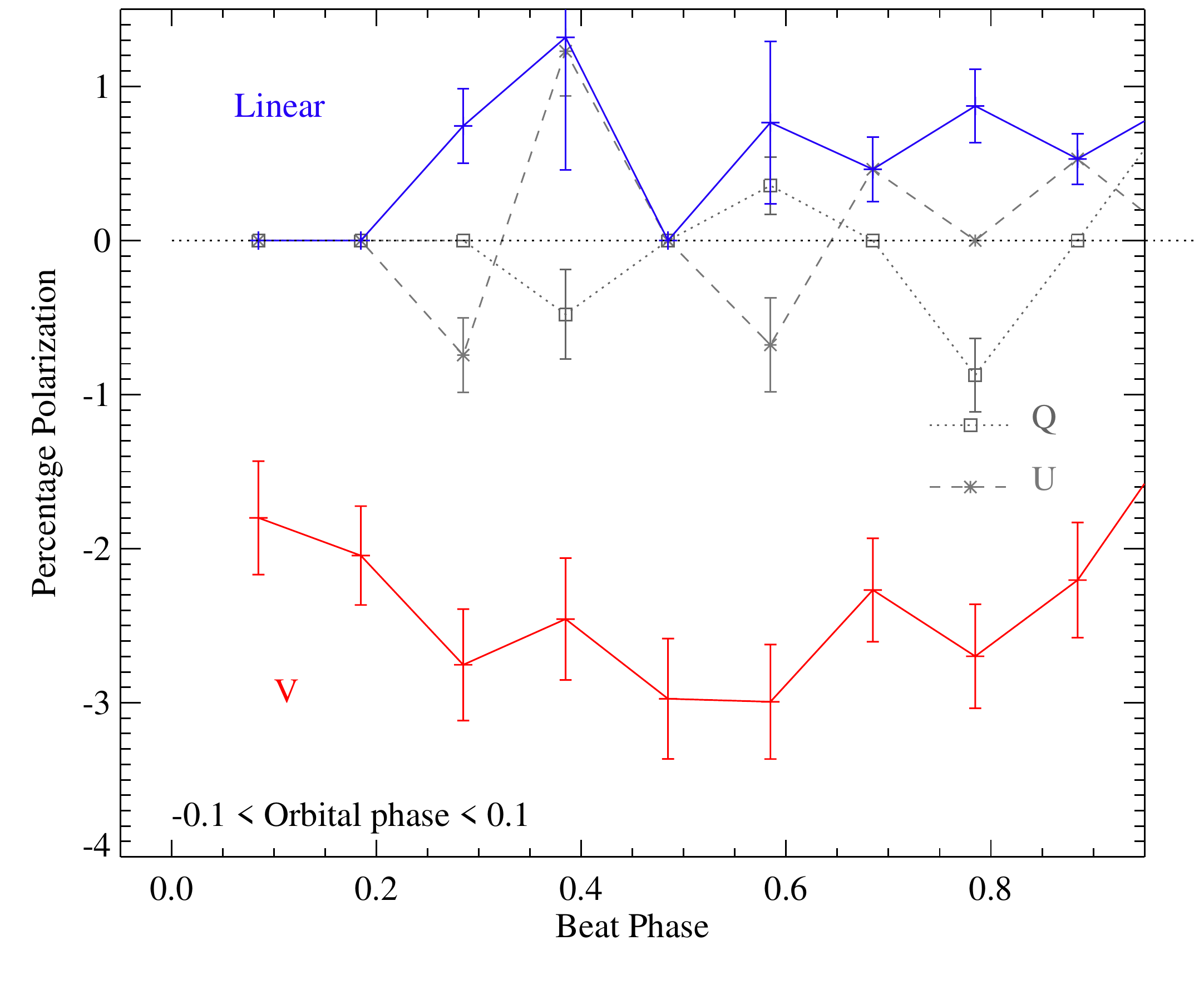}
\includegraphics[width=0.75\columnwidth]{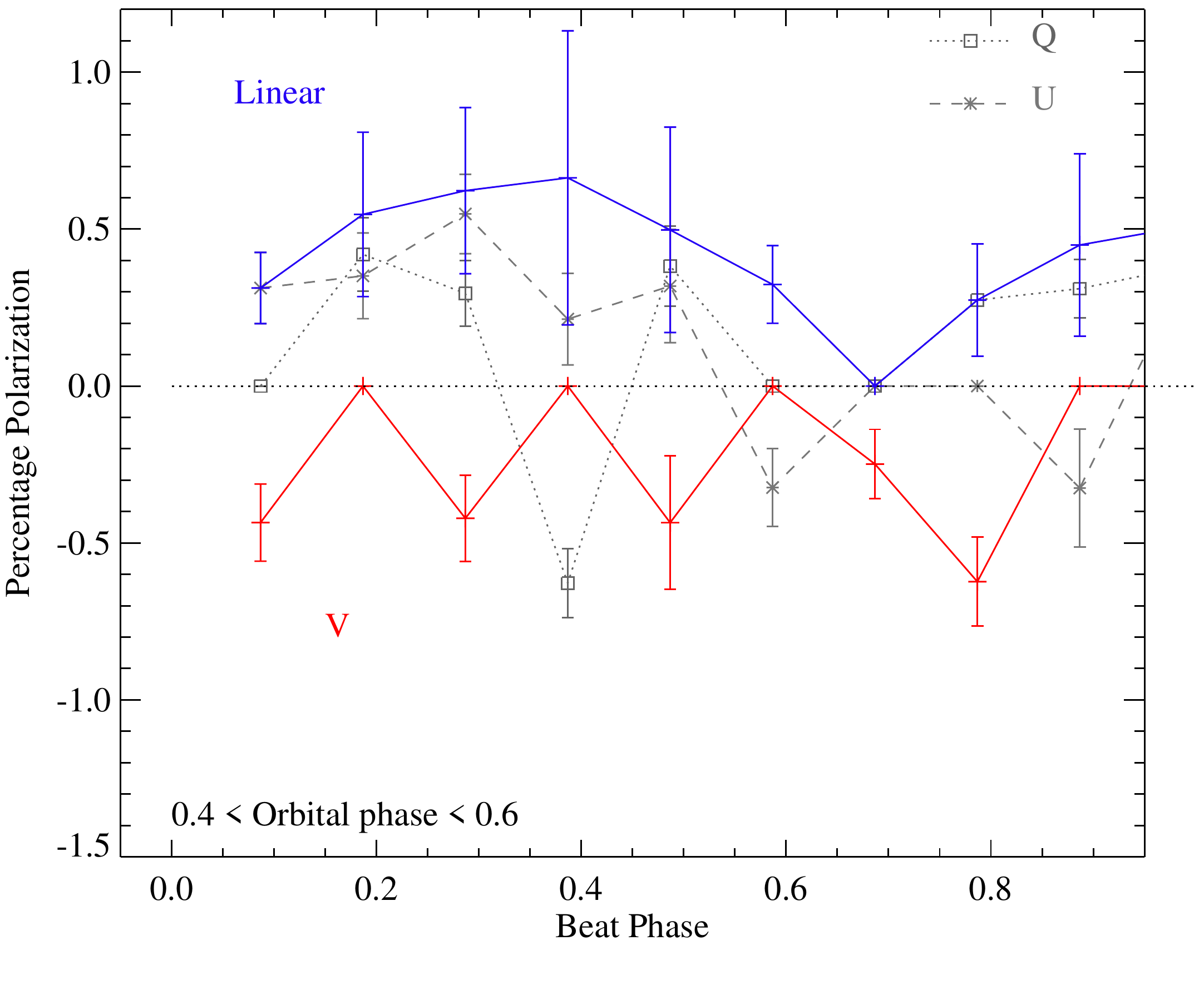}
\caption{The dependence of polarization on orbital phase and beat phase at 5\,GHz. As in figure \ref{fig:pol_beats9}.}
           \label{fig:pol_beats5}%
\end{figure}

\begin{figure}
\centering
\includegraphics[width=0.75\columnwidth]{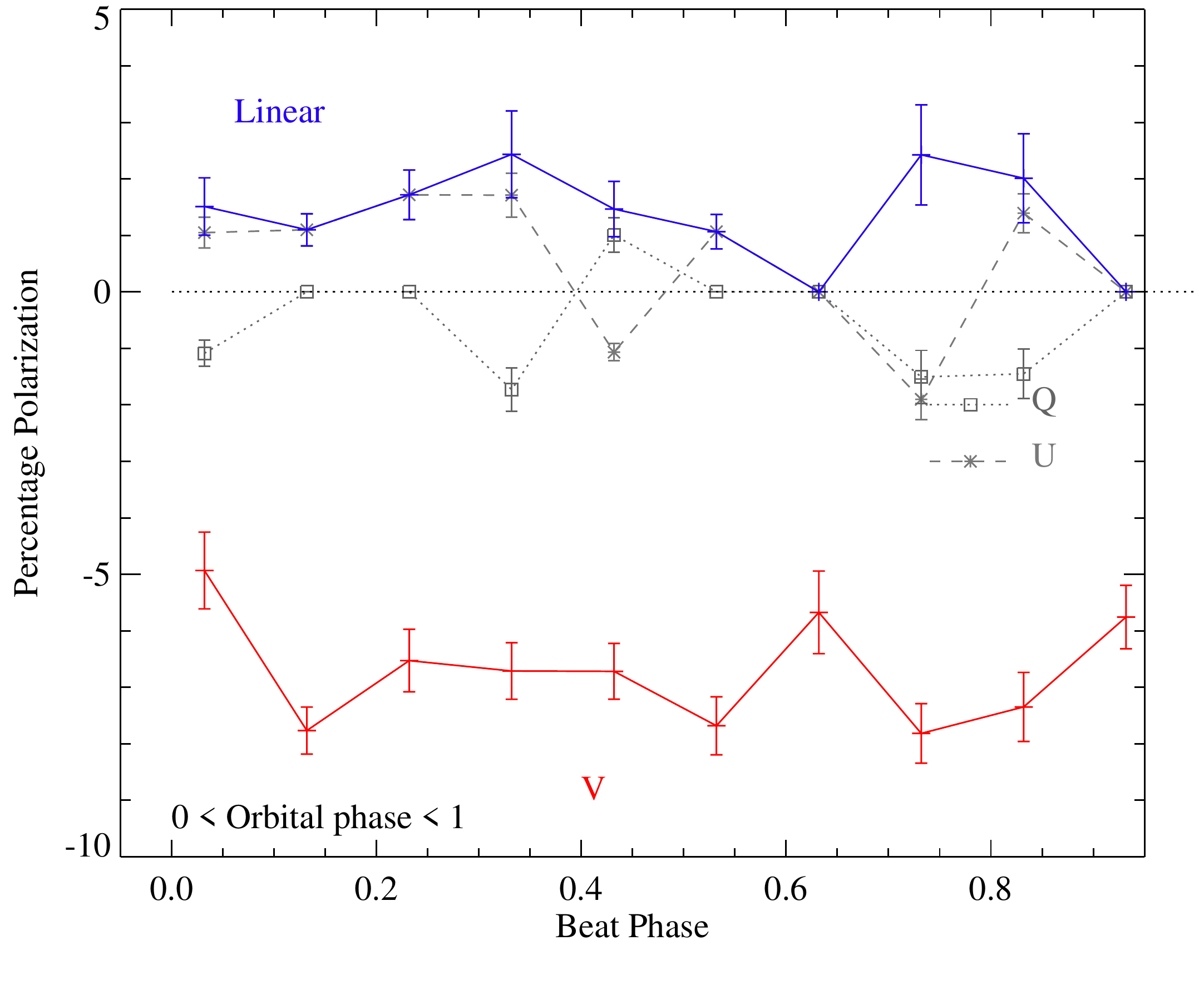}
\includegraphics[width=0.75\columnwidth]{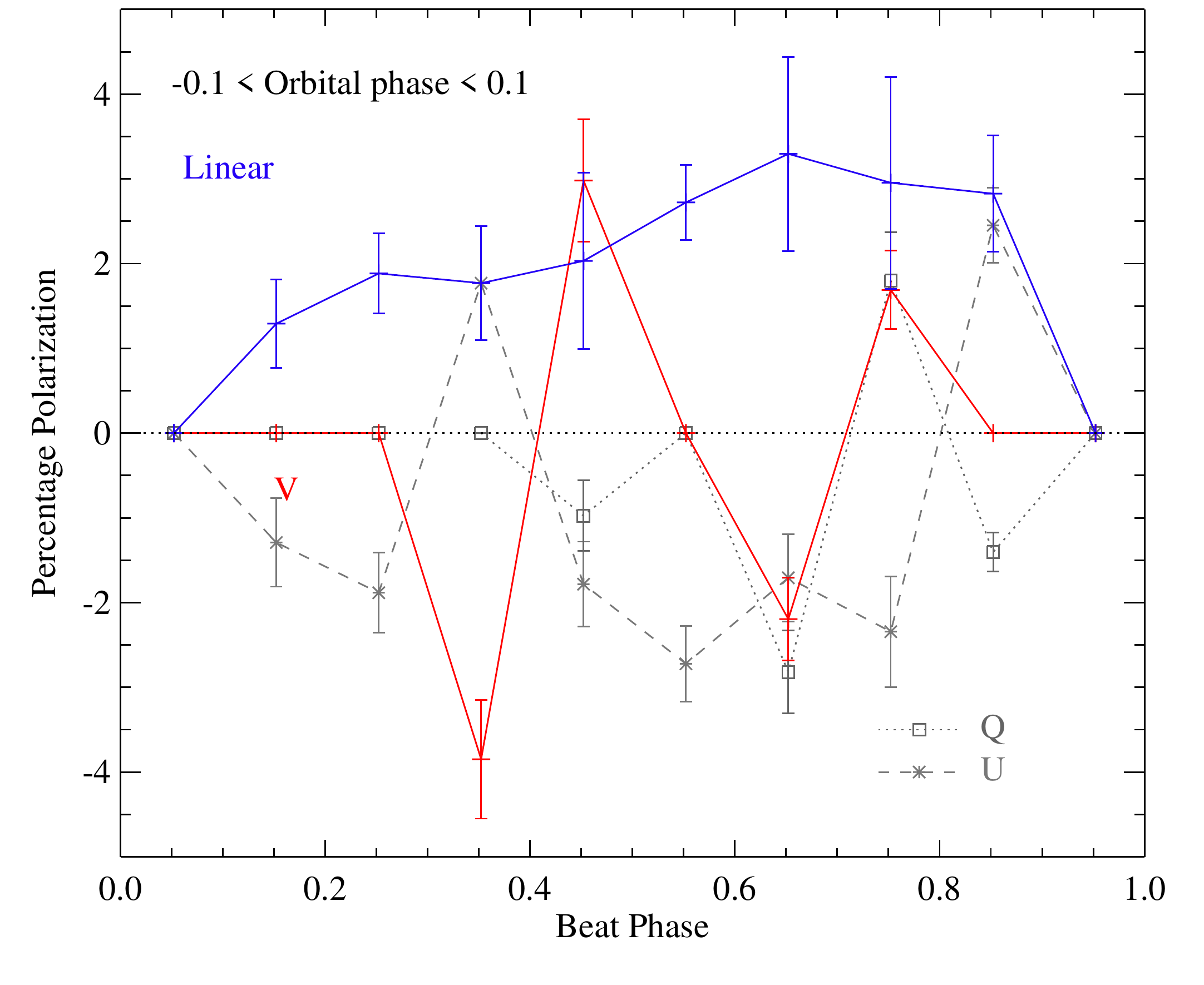}
\includegraphics[width=0.75\columnwidth]{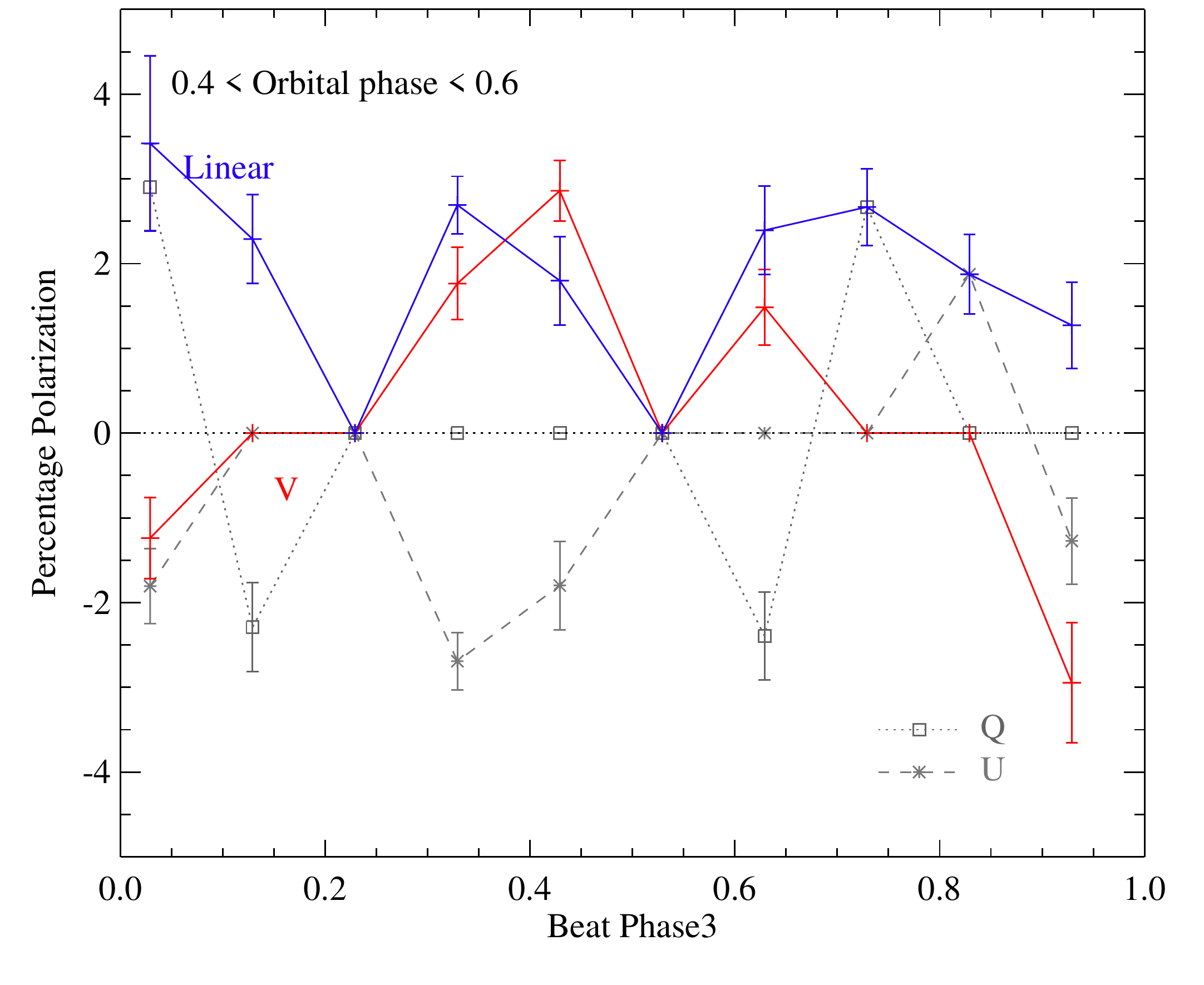}
\caption{The dependence of polarization on orbital phase and beat phase at 1.5\,GHz. As in Figure \ref{fig:pol_beats9}.}
           \label{fig:pol_beats1}%
\end{figure}

\subsection{Polarization with Beat Phase}

\subsubsection{9\,GHz band}
In optical light, the polarization of \arsco\ is known to vary strongly on the beat and spin periods \citep{2017NatAs...1E..29B}. In the radio, we lack the signal-to-noise and $uv$-plane coverage to resolve the polarization on these two minute periods. Instead we investigate behaviour using phase folding as before. Individual visibility integrations were assigned a beat phase, based on the 118.2\,s beat period and referenced to the same (5\,GHz beat minimum) zero phase as in section \ref{sec:beat}. The visibilities were then grouped by this beat phase, and an image created in each Stokes parameter, including only those visibilities contributing to a given beat phase bin. This was done for the entire observation (i.e. a full orbital period) and separately for subsets of the data covering 20\% in orbital phase, centred near the maximum and minimum of the orbital flux variation.  The resulting polarization light curves are shown in Figure \ref{fig:pol_beats9}.

As is clear from the figures, the very low linear polarization at 9\,GHz, in contrast to the 40\% polarization seen in the optical, does not arise entirely from the effects of averaging over the beat cycle. Splitting the data by beat phase suggests that the linear polarization is nearly constant through the beat cycle, at a low level with no clear dependence on beat phase. There is also no clear dependence of the linear polarization beat-period lightcurve on orbital phase. The mean linear polarization in the upper panel of Figure \ref{fig:pol_beats9} (i.e. dividing by beat phase for the whole orbital period) is 0.16$\pm$0.03\%. However it is possible that this is averaging out some of the signal. When the data for the two subsets in the lower panels of Figure \ref{fig:pol_beats9} are considered, the mean 9\,GHz linear polarization increases to 0.37$\pm$0.04\%. This suggests that there are short intervals with significantly higher polarization than the mean. If the polarization angle varies (i.e. signal is exchanged between the Stokes $Q$ and $U$ parameters), such short-term polarization signals would be washed out in a time-averaged flux. There is some hint of this happening - while the total linear polarization remains broadly constant through the beat period, both $Q$ and $U$ fluxes vary significantly in adjacent phase bins.

By contrast, the 9\,GHz circular polarization, shown in Figure \ref{fig:polorbit9} to be strongly dependent on orbital phase, also appears to be sensitive to beat phase. The Stokes $V$ parameter shows negative flux (i.e. counterclockwise circular polarization in the signal), at all beat phases when averaging over the orbit, or when considering the minimum of the Stokes $I$ lightcurve. It is strongest at a  beat phase of 0.5 (where zero marks the minimum total flux), peaking at $V=-4.2\pm0.3$\%.

Finally, there is no clear evidence for polarization variation when folding on the spin (rather than beat) frequency. Reproducing Figure \ref{fig:pol_beats9} on the spin period yields near-constant linear polarization at the same level ($\sim0.4$\%), while circular polarization shows a similar strength and pattern to that folded on the beat period (i.e. appearing strongest, with $V\sim$-4\%, at the minimum of the folded light curve).

\subsubsection{5\,GHz band}

As before, we repeat our analysis at 5\,GHz, showing the results in Figure \ref{fig:pol_beats5}. Beat period polarization modulation at this frequency mirrors that at 9\,GHz. The linear polarization does not exceed 1\% at any beat phase (when phase-folded) but is typically non-zero. The circular polarization is again modulated on the beat-phase, peaking at a beat phase of 0.4-0.5.

\subsubsection{1.5\,GHz band}

\begin{figure}
\centering
\includegraphics[width=1\columnwidth]{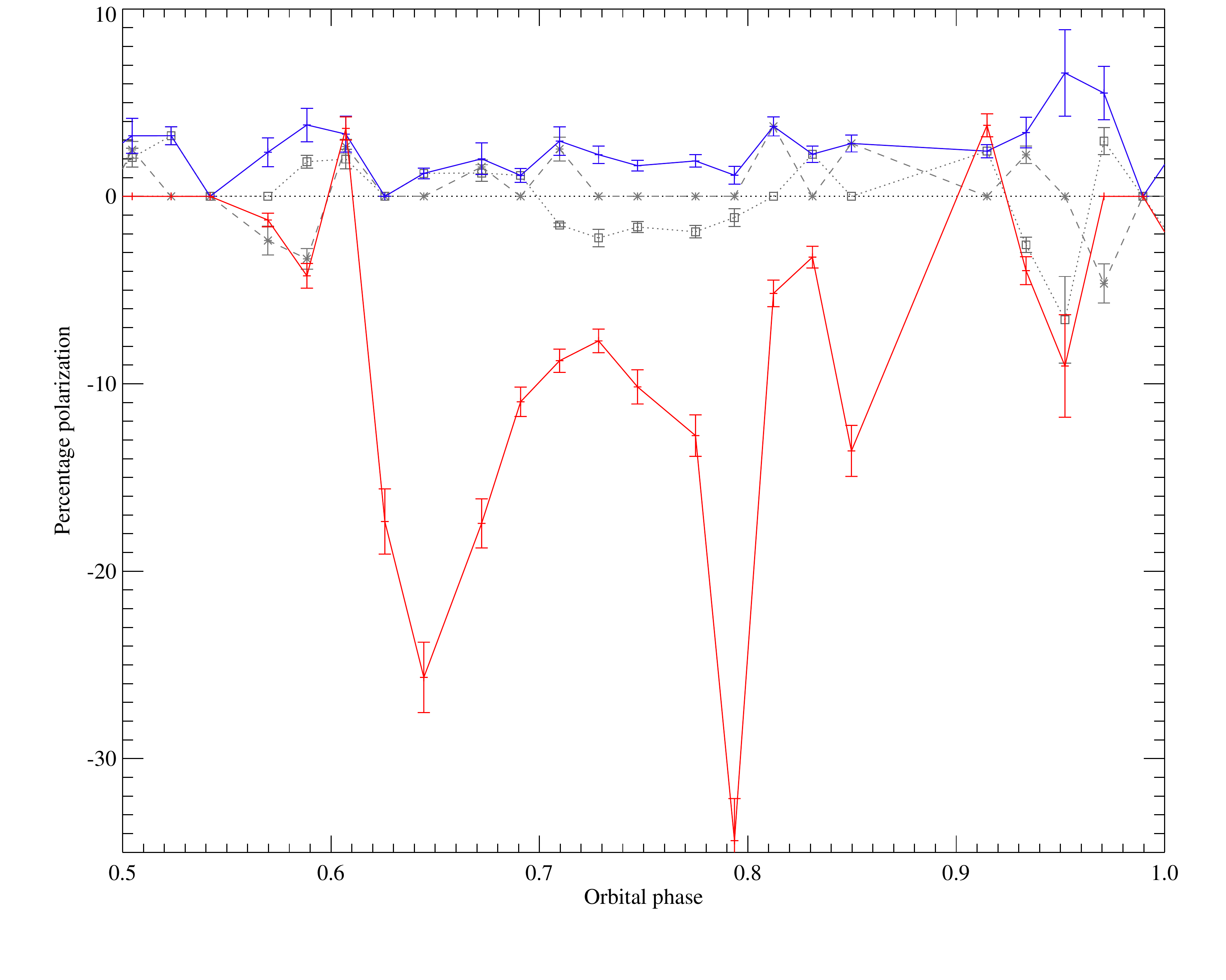}
\includegraphics[width=1\columnwidth]{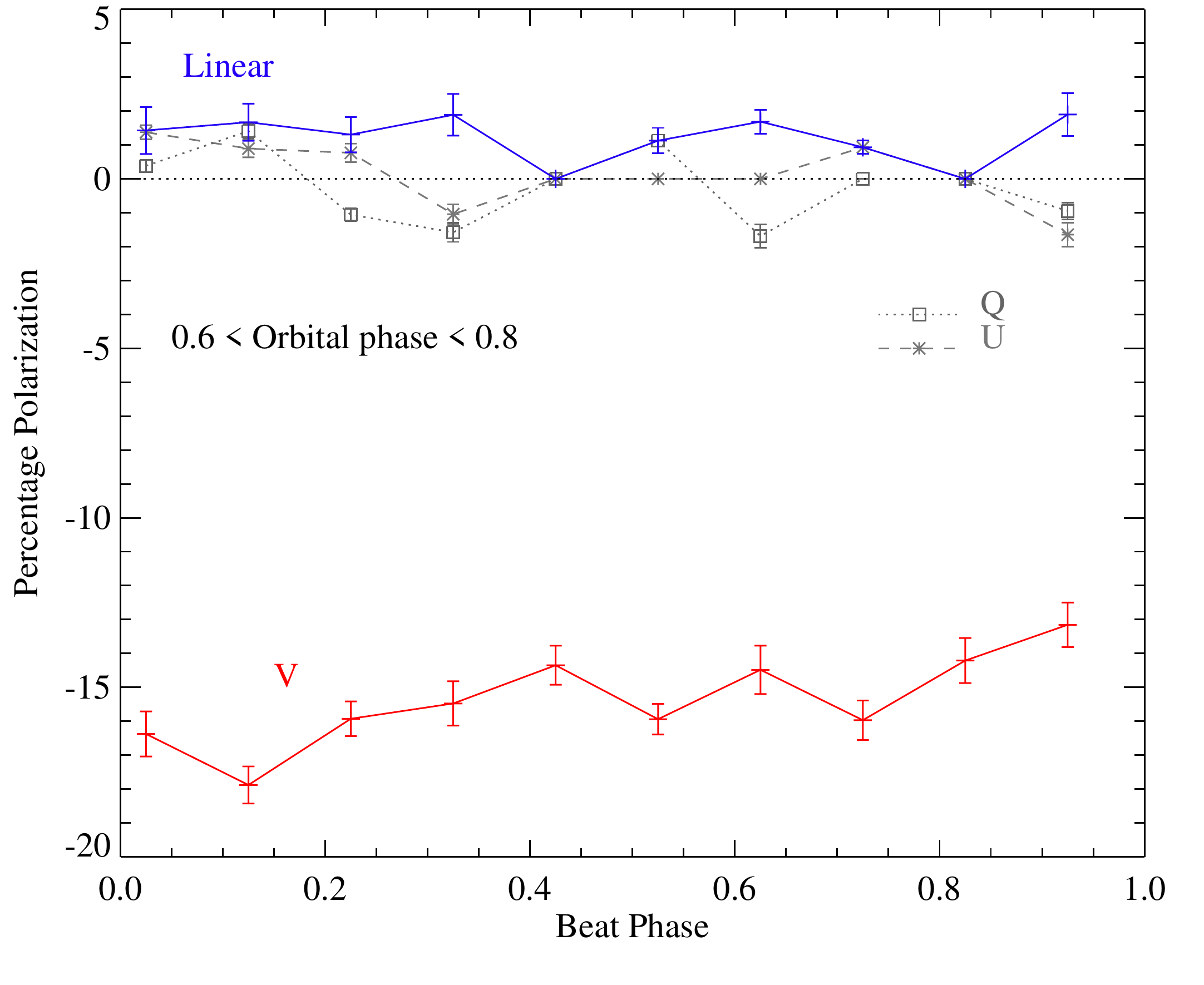}
\caption{Closer study of the negative circular polarization excursion of the 1.5\,GHz orbital lightcurve, now imaged in 4 minute intervals, rather than 10 minutes. In the lower panel we show the dependence of polarization on beat phase during this excursion.}
           \label{fig:circzoom}%
\end{figure}

The results of folding the 1.5\,GHz data on the beat period are shown in Figure  \ref{fig:pol_beats1}.  These data show less evidence for a beat period dependence in the polarization. Given the lower signal-to-noise ratio in the total flux measurements, uncertainties on the polarization are larger in this band. Both the linear and circular polarization show near-constant values, at $\approx$1.5 and $-7$\% respectively. While there is some variation with beat phase, this is consistent with the uncertainties on each data point. At this frequency, orbital phases of 0 and 0.5 both miss significant polarized emission features in the orbital light curve, and the beat-folded light curve shows no clear evidence for a beat-phase modulation in 1\,GHz data near these phases, albeit with large uncertainties on individual data points. We also consider the beat-cycle at orbital phase 0.4 - 0.6 (where the Stokes V flux is large and negative) in figure \ref{fig:circzoom}. As expected, the circular polarization is larger than the orbit average, but still shows no clear dependence on beat phase.

%-------------------------------------------------------------------

\section{Discussion and Interpretation}\label{sec:disc}

\begin{figure}
\centering
\includegraphics[width=1\columnwidth]{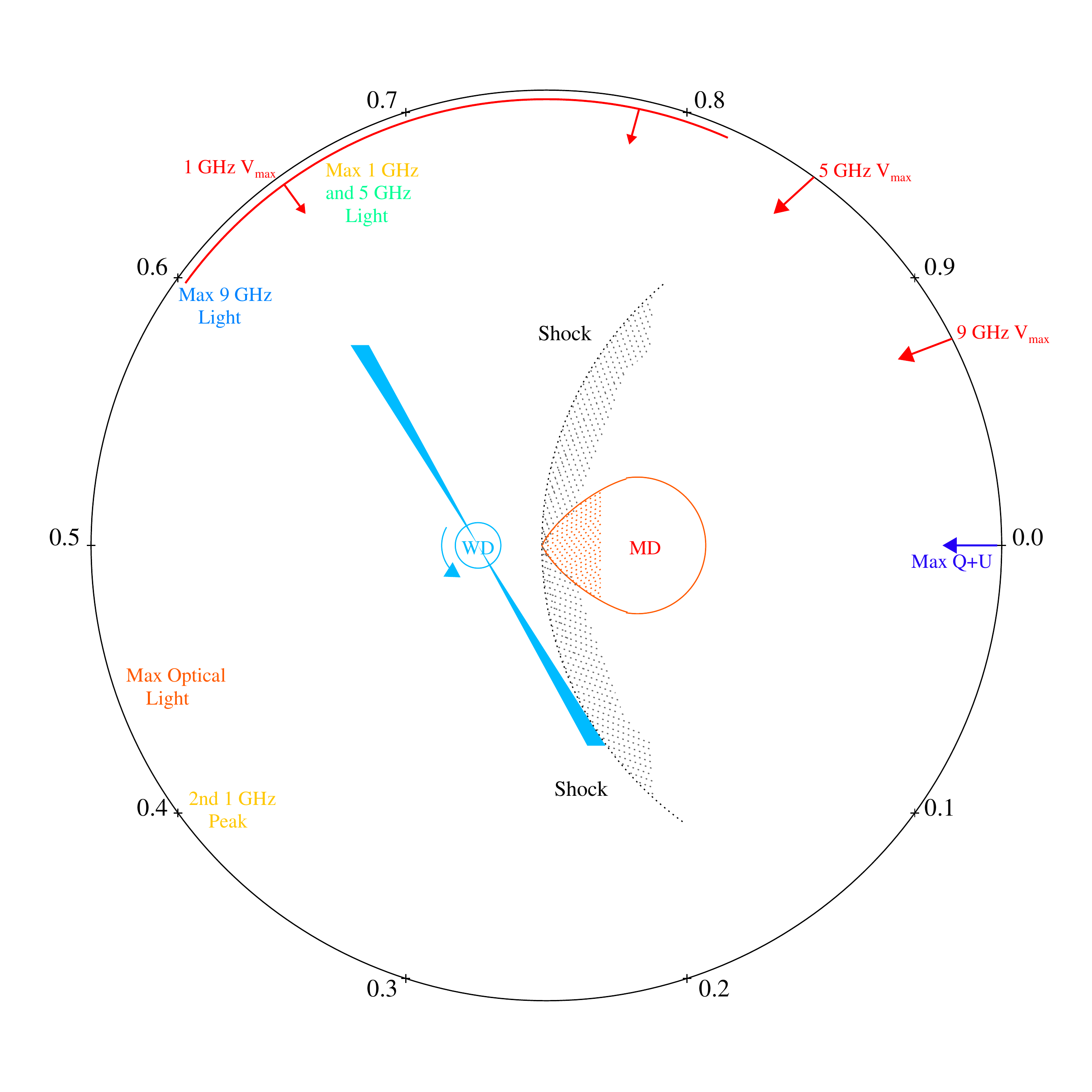}
\caption{Orbital phase diagram for the \arsco\ system (not to scale). Key phases at which the system output peaks are indicated. We also show the beamed emission proposed for the white dwarf and potential shocks where it meets the white dwarf magnetosphere.}
           \label{fig:diagram}%
\end{figure}

The strong periodic variability of \arsco\ in the optical marked it out as an anomalous source among known white dwarf binaries. Our initial radio observations, obtained with ATCA, demonstrated that these extended to the radio. However they were too short to probe the orbital phase behaviour, and too limited by $uv$-plane coverage and sensitivity to probe short time-scales at high signal to noise.

Here we have demonstrated that the GHz radio emission of \arsco\ is in many ways similar to that observed in the optical. The orbital phase modulation of the total continuum flux is clearly detected across the full range of observed frequencies (1-10\,GHz), with the orbital modulation exceeding 50\% of the peak flux. The minimum of the radio lightcurve is also at a  similar orbital phase to that observed in the optical, albeit with evidence for radio frequency-dependent variation in the light curve. This provides strong evidence that the radio continuum emission originates from close to the inner face of the M-dwarf, if not from the stellar surface itself. Key orbital phases for the \arsco\ system are indicated on a phase diagram for clarity in figure \ref{fig:diagram}.

The radio spectral slope $S_\nu \propto \nu^{0.35}$ is comparable to the self-absorbed synchrotron slope of  $S_\nu \propto \nu^{0.33}$ derived by \citet{2016ApJ...831L..10G} for the 10-1000\,GHz region, from data in \citet{2016Natur.537..374M}, but significantly shallower than the self-absorbed spectral slope, $\nu^2$, they predict for the 1-10\,GHz region. Where the total flux peaks, at orbital phase of 0.5, the spectral index steepens. This may suggest that an additional emission component is contributing to the continuum emission at low radio frequencies ($<5$\,GHz), with the strongest effect where the synchrotron emission spectrum that extends up to the optical is weakest. The orbital light curves shown in Figure \ref{fig:orbital_flux} suggest that this low-frequency emission component peaks twice in the orbital phase, which may imply a bipolar emission source arising from, or near, the (tidally-locked) M-dwarf.

The radio data also confirms strong modulation of the total flux at 5 and 9\,GHz on the system beat period and at half this period,
with little signal on the spin period. Again the double-peaked light curve suggests bipolar emission, this time coming from an interaction between the white dwarf and red dwarf, consistent with the extant picture of this source as a nearly-perpendicular rotator, with the white dwarf's polar magnetic fields whipping past the red dwarf twice in each white dwarf spin period \citep{2016Natur.537..374M}. The very erratic flaring and variation in strength of the signal from pulse to pulse suggests short timescales (a matter of seconds) for variation in the radio emission, suggesting the flaring region is small. 

Interestingly these modulations disappear at the lowest frequencies, with the 1.5\,GHz data showing no significant evidence for modulation on either the beat or spin frequencies. This hints that the 1.5\,GHz flux may arise from a distinct, non-pulsing mechanism in \arsco, as suggested by the double-peaked orbital lightcurve and the shallow spectral slope.

The exceptional linear polarization seen in the optical is not seen at radio frequencies. The linear polarization of \arsco\ in the radio is near constant at $<1$\%. However, the presence of well-detected circular polarization implies the presence of magnetic fields aligned along the line of sight, and its modulation on both the orbital and beat periods requires further consideration. To first order, the ratio between circular and linear polarization is determined by the Lorentz factor of the emitting electrons, with strong circular polarization implying non-relativistic, cyclotron rather than relativistic synchrotron emission.

This emission source appears to be near-constant, with the beating but unpolarized (in the radio) synchrotron emission superimposed upon it. If so, the polarized emission source likely has a circular polarization well in excess of the measured 5-10\% seen at 9\,GHz and is simply being diluted by the stronger, pulsed emission. This is supported by the fact that polarization is also stronger at lower frequencies where the total flux is lower.

Interestingly \arsco\ shows very similar orbital phase polarization behaviour to known polars and intermediate polars (IPs) - a class of White Dwarf-M dwarf binaries with which \arsco\ shares many of its properties. These sources have typical magnetic fields $<$100\,MG \citep{2015SSRv..191..111F}, 
slightly lower than that inferred for emission regions in \arsco\ from optical polarization \citep{2017NatAs...1E..29B}, and are often characterised by cyclotron emission features in the optical. Several known polars, including VV~Puppis \citep{1982MNRAS.198..975W}, EXO 03319-2554.2 \citep{1989ApJ...337..832F} and MLS110213 \citep{2015MNRAS.451.4183S}, show optical circular polarization peaking at about 10\% of the total flux. 
In IPs, polarization is also observed to modulate on the white dwarf spin period \citep[e.g.][]{2012MNRAS.420.2596P} while sources typically show linear polarization comparable to circular, with $L/V \sim 0.2-0.6$ \citep{2000PASP..112..873W}, comparable to that in \arsco\ ($L/V \sim 0.2$ at 5\,GHz). In the case of both polars and IPs, this is interpreted as a cyclotron emission region arising from single-pole stream-fed accretion onto the surface of the white dwarf.

In \arsco, despite the striking similarity in the circularly polarized lightcurve, and the circular-to-linear polarization ratio, this model cannot apply in all particulars. There is no sign of accretion, suggesting that the white dwarf cannot contribute in the same manner as in IPs. The much weaker variation on the spin or beat periods, relative to that observed at near-zero orbital phase, also suggests that the circularly-polarized emission arises primarily from the M dwarf itself, perhaps from steady accretion of magnetically-entrained winds onto polar regions of the red dwarf or from emission in a bow shock region near the surface.

While the white dwarf in \arsco\ has a strong magnetic field, the environs of the M dwarf are likely to experience lower field strengths. Their typical surface fields are in the region of 0.1 to 1kG (although they can be stronger in localised regions) \citep[e.g.][]{2006ApJ...648..629B}. This is comparable to the estimated strength of the white dwarf field at the M dwarf surface \citep[0.1\,kG,][]{2017ApJ...835..150K}. These field strengths correspond to non-relativistic cyclotron  emission frequencies from 0.3 to 3\,GHz, which would place the fundamental emission peak for cyclotron flux from the M dwarf within, or just below, our observation frequencies. Our radio observations of \arsco\ are thus analogous to optical observations of polars in the sense that the cyclotron frequency of polars is usually in the near-infrared, so that harmonics of order a few are observed in the optical, while we may be seeing the equivalent low order cyclotron / gyro-synchrotron emission from the M dwarf in \arsco\ in the radio.

This emission may be most visible when the irradiated face of the red dwarf (which pulses with strong synchrotron emission on the beat period) is turned out of the line of sight and no longer swamps the lower energy cyclotron emission. 
As well as the different cyclotron emission site (M dwarf, rather than white dwarf),
\arsco's broad synchrotron spectrum shows that it is distinguished from typical polars by the presence of an extended high energy tail of electrons. These relativistic electrons generate the higher frequency radio, far infrared, infrared, optical and UV emission. If this comes from a region with a $\sim$100\,G field, this emission implies electrons with Lorentz factors of order $\sim$1000, or energies of order a GeV, so it is true synchrotron (rather than cyclotron emission) and we should not expect such high circular polarisation. This high energy tail might act to wash out the cyclotron hump structure that one might otherwise expect at radio frequencies if the analogy with optical observations of polars were exact.

%-------------------------------------------------------------------

\section{Conclusions}\label{sec:conc}

We have obtained high time resolution, interferometric radio  spectral imaging and polarization data for AR Sco in the 1.5, 5 and 9\,GHz bands, using the Karl G. Jansky Very Large Array.  Our main conclusions can be summarized as follows:

\begin{itemize}
\item  The total radio flux from \arsco\ exhibits a power law spectrum between 1 and 10\,GHz, with a spectral slope of $0.358\pm0.015$ (fit in the range 4-10\,GHz) when flux is integrated over a full orbit.

\item  The total radio flux is modulated on the 3.56\,hr orbital period of the \arsco\ binary system. The strength of this modulation is a function of frequency, with the strongest changes seen at 9\,GHz.

\item  Emission at 5 and 9\,GHz is also modulated on the white dwarf spin-orbit beat period also identified in optical data. Emission at 1.5\,GHz shows no apparent beat phase modulation. The fractional pulsation strength decreases with frequency, and is lower in the radio than the optical. The median beat pulse fraction is $\sim$20\% at 9\,GHz.

\item  \arsco\ shows significant levels of negative circular polarization in a narrow range of orbital phases. The peak of this circularly polarized emission shifts to earlier phases with decreasing frequency, as the fractional strength of the circularly polarized flux increases. Circularly polarized emission also peaks at the minimum of the beat-folded light curve at 5 and 9\,GHz.

\item The linear polarization of \arsco\ is far lower in the radio than optical, peaking at a few percent at 1.5,GHz, and shows no clear dependence on beat phase at any frequency.

\item Emission at 1-10\,GHz likely arises from close to or on the surface of the M-dwarf, with no clear evidence for emission from the white dwarf.

\item The 1-10\,GHz frequency regime appears to mark a transition between different dominant emission regions and mechanisms, likely breaking from the synchrotron power-law emission that extends up to the optical towards a non-relativistic, circularly polarized cyclotron emission mechanism at low frequencies.

  \end{itemize}

In conclusion, the radio observations of \arsco\ have revealed a hitherto unseen, non-relativistic electron emission component at low radio frequencies. The system nonetheless remains a challenging source to interpret, and would benefit from further observations and theoretical study. Simultaneous observations at multiple frequencies in the radio may be necessary to fully disentangle flaring and orbit-to-orbit variation from frequency dependent shifts in behaviour. This presents significant technical challenges, both in terms of the shortage of arrays with multiband receivers, and the difficulty of obtaining sufficient signal to noise and support in the $uv$-plane for the short integrations required.

%-------------------------------------------------------------------

\begin{acknowledgements}

TRM, PJW and DS are supported by UK Science and Technology Facilities Council (STFC) Consolidated Grant ST/P000495/1.  
The research leading to these results has received funding from the European Research Council under the European Union's Seventh Framework Programme (FP/2007-2013) / ERC Grant Agreement n. 320964 (WDTracer).

The National Radio Astronomy Observatory is a facility of the National Science Foundation operated under cooperative agreement by Associated Universities, Inc.
  
\end{acknowledgements}

%------------------------------------------------------------------

\end{document}